\documentclass[%
reprint,
 amsmath,amssymb,
aps,
floatfix,
]{revtex4-2}

\usepackage{hyperref} 
\usepackage{siunitx}
\usepackage{commath} 
\usepackage{bbm}
\usepackage{amsfonts}
\usepackage[section]{placeins} 
\usepackage{graphicx}
\usepackage{dcolumn}
\usepackage{bm}
\usepackage{float}
\usepackage[normalem]{ulem} 
\usepackage{xcolor} 
\usepackage{inputenc}

\begin{document}


\title{Chiral quantum optics in broken-symmetry and topological photonic crystal waveguides}

\author{Nils Valentin Hauff}
 \affiliation{Center for Hybrid Quantum Networks (Hy-Q), Niels Bohr Institute, University of Copenhagen, Blegdamsvej 17, DK-2100 Copenhagen, Denmark}

\author{Hanna Le Jeannic}
 \altaffiliation[Present address: ]{Laboratoire Photonique Numérique et
Nanoscience, Université de Bordeaux, Institut d’Optique, CNRS,
UMR 5298, 33400 Talence, France}
\affiliation{Center for Hybrid Quantum Networks (Hy-Q), Niels Bohr Institute, University of Copenhagen, Blegdamsvej 17, DK-2100 Copenhagen, Denmark}%

\author{Stephen Hughes}
\affiliation{Centre for Nanophotonics, Department of Physics, Engineering Physics \& Astronomy, 64 Bader Lane, Queen’s University, Kingston, Ontario, Canada K7L 3N6}%

\author{Peter Lodahl}
\affiliation{Center for Hybrid Quantum Networks (Hy-Q), Niels Bohr Institute, University of Copenhagen, Blegdamsvej 17, DK-2100 Copenhagen, Denmark}

\author{Nir Rotenberg}
\affiliation{Center for Hybrid Quantum Networks (Hy-Q), Niels Bohr Institute, University of Copenhagen, Blegdamsvej 17, DK-2100 Copenhagen, Denmark}%
\affiliation{Centre for Nanophotonics, Department of Physics, Engineering Physics \& Astronomy, 64 Bader Lane, Queen’s University, Kingston, Ontario, Canada K7L 3N6}%

\date{\today}

\begin{abstract}
On-chip chiral quantum light-matter interfaces, which support directional interactions, provide a promising platform for efficient spin-photon coupling, non-reciprocal photonic elements, and quantum logic architectures. We present full-wave three-dimensional  calculations to quantify the performance of conventional and topological photonic crystal waveguides as chiral emitter-photon interfaces. Specifically, the ability of these structures to support and enhance directional interactions while suppressing subsequent backscattering losses is quantified. Broken symmetry waveguides, such as the non-topological glide-plane waveguide and topological bearded interface waveguide are found to act as efficient chiral interfaces, with the topological waveguide modes allowing for operation at significantly higher Purcell enhancement factors. Finally, although all structures suffer from backscattering losses due to fabrication imperfections, these are found to be smaller at high enhancement factors for the topological waveguide. These reduced losses occur because the optical mode is pushed away from the air-dielectric interfaces where  scattering occurs, and not because of any topological protection. These results 
are important to the understanding of light-matter interactions in topological photonic crystal and to the 
design of efficient, on-chip chiral quantum devices.
\end{abstract}


\maketitle



\section{\label{sec:Intro}Introduction}


Chiral light-matter interactions are directional, enabling non-reciprocal devices and circuits, including unidirectional single photon emission.
This directionality arises from the interaction of elliptical dipoles  with finely structured light fields of nanophotonic systems such as plasmonic surfaces \cite{rodriguez2013near,lin2013polarization}, nanowaveguides \cite{petersen2014chiral,coles2017path}, resonators \cite{schneeweiss2017fiber, martin2019chiral} and photonic-crystal waveguides (PhCWs) \cite{le2015nanophotonic,sollner2015deterministic,PhysRevLett.115.153901} using either classical or quantum light sources \cite{lodahl2015interfacing}. Chiral light-matter interactions enable non-reciprocal devices such as optical isolators \cite{sollner2015deterministic}, circulators \cite{xia2014reversible,sayrin2015nanophotonic,scheucher2016quantum} and quantum gates \cite{shomroni2014all,pedersen2021demand} and, in waveguides, are the basis for several protocols for quantum networks \cite{mahmoodian2016quantum}.

Recently, a new class of topologically engineered PhCWs have been proposed and demonstrated \cite{khanikaev2013photonic,raghu2008analogs,ma2016all,Hafezi_2011,wang2009observation,wang2009observation,PhysRevLett.100.013904,PhysRevLett.113.113904}, generating considerable excitement within the quantum optics community \cite{mehrabad2020chiral,yamaguchi2019gaas,barik2018topological,li2009topological}. In analogy to edge modes of electronic topological insulators \cite{RevModPhys.83.1057,RevModPhys.82.3045}, the guided modes of topological PhCWs have been identified as both \textit{chiral} \cite{Hafezi_2011,he2019silicon,fang2012realizing} and resistant to backscattering in 60 degree bends \cite{PhysRevLett.126.027403,yang2018topological,yamaguchi2019gaas,shalaev2019robust,mehrabad2020chiral}. 
Light propagation in all-dielectric media is described by Maxwell's equations obeying time-reversal symmetry (TRS) demanding that the complex electric field in the forward direction is the complex conjugate of the electric field in the backward direction. Light scattering from inhomogeneities (classical or quantum) alters the flow of light. In a one-dimensional waveguide, the condition referred to as chiral interaction corresponds to forward and backward propagating modes scattering with different strengths from the same (elliptically polarized) dipole. In this situation, a radiating dipole will be directional, which is referred to as chiral emission. Chiral single-photon emission has been demonstrated in topological PhCWs \cite{barik2018topological, mehrabad2020chiral,yamaguchi2019gaas}. However, only few works have considered scattering losses beyond 60 degree bends that constitute inherent symmetry directions of the structure \cite{shalaev2019optically,blanco2018topological,jin2019topologically,arregui2020quantifying}. Potentially topological waveguides could have increased robustness towards backscattering loss from inherent fabrication imperfections in real devices. This could be of technological importance in quantum photonic devices where inherent backscattering leads to Anderson localization of light  \cite{sapienza2010cavity,crane2017anderson,PhysRevB.95.224202} thus limiting the operation length of the devices.
Disorder-induced scattering also places a limit on 
device length for exploiting nonlinear
effects in slow light PhCWs~\cite{PhysRevLett.118.253901,husko2016free}.

In this work, we use full vectorial three-dimensional finite-element simulations and rigorous scattering theory to explore how well different photonic crystal waveguides (Sec. II), and more specifically their guided modes (Sec. III), can act as quantum chiral-light matter interfaces (Sec. IV). We find that topological waveguides can be favorable compared to standard line-defect (W1) and even glide-plane waveguides \cite{mahmoodian2017engineering} (GPWs) as they enable near-unity directionality at higher Purcell factors and lower propagation losses (Sec. V). We stress that the differences in loss performance  arise from variations of the electric field distribution rather than the topological nature of the edge state modes, here based on the Quantum Spin-Hall effect \cite{raghu2008analogs}.
Finally, we show that topological waveguides, despite this lack of topological protection \cite{PhysRevB.101.054307}, can outperform conventional photonic crystal waveguide for realization of integrated non-reciprocal single-photon devices 
for constructing of scalable complex quantum circuits and networks \cite{lodahl2017chiral}  (Sec.~VI).

\section{Photonic crystal waveguides as quantum chiral interfaces}\label{sec:ChiralInterface}
\begin{figure}
    \centering\includegraphics[ scale=1.0]{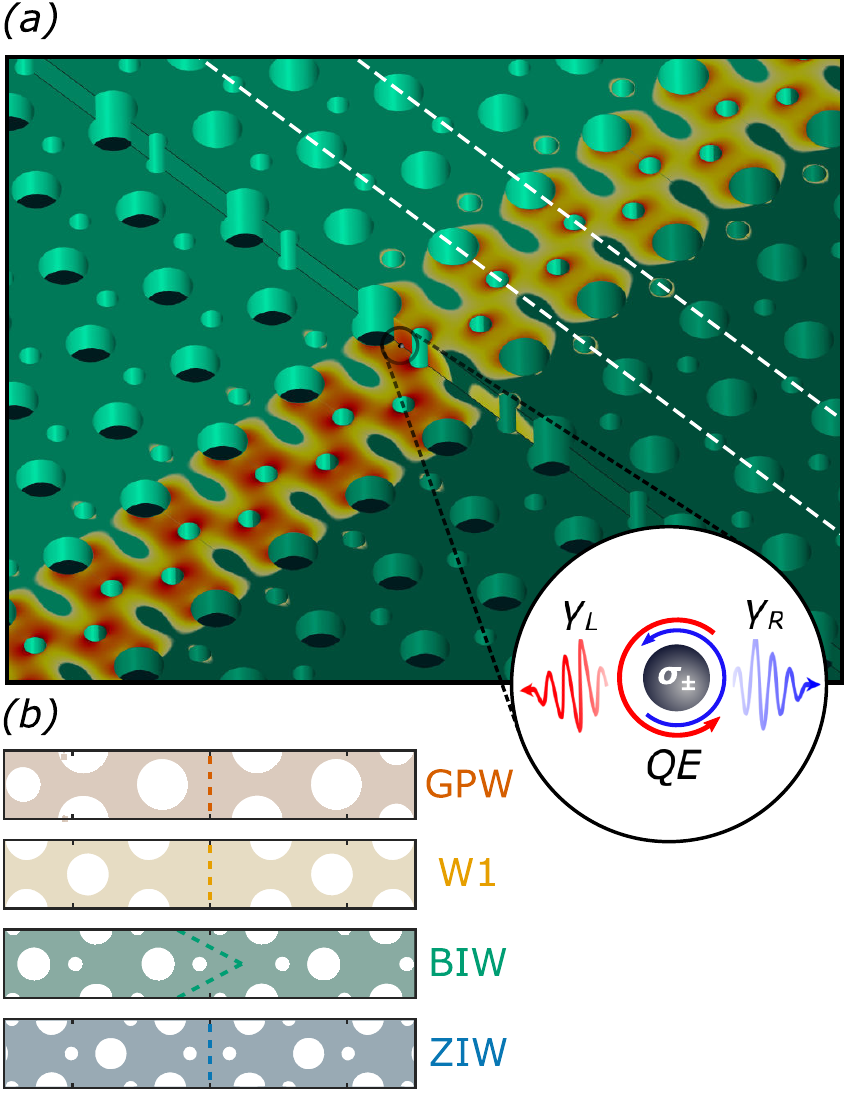}
    \caption{ (a) Schematic of a chiral light-matter interface utilizing an edge mode between two topological photonic insulators (dark and bright green). The field norm %
    for the guided edge-mode is shown (yellow to red) and an embedded quantum emitter indicated. The quantum emitter's transition dipoles are left- or right-handed circular ($\sigma_{\pm}$), resulting in directional emission (left and right as shown in the inset) when it is placed at a position where the polarization of the guided mode is circular. The white-dashed line highlights a single supercell of the waveguide. (b) Schematics of the supercells of the BIW, ZIW, GPW, and W1 with the interface or center of each waveguides highlighted with a dashed line. %
    }
    \label{fig:Schematic}
\end{figure}

An ideal chiral interface for quantum light-matter interactions, such as the one shown in Fig.~\ref{fig:Schematic}, is characterized by several properties. First and foremost, emission or scattering of photons by a quantum emitter (QE)  into the counter-propagating modes left (L) or right (R) should be highly asymmetric with decay rates $\gamma_{\mathrm{L}} \gg \gamma_{\mathrm{R}}$, or vice versa. This directionality occurs when the overlap of circular (or elliptical) transition dipoles with the two counter-propagating modes differs, and is quantified by,
\begin{equation}\label{eq:D}
    D = \frac{\gamma_{\mathrm{L}}-\gamma_\mathrm{R}}{\gamma_\mathrm{L}+\gamma_\mathrm{R}}.
\end{equation}
Below, we show how this directionality factor can be calculated for any electric field profile.

Second, an efficient quantum light-matter interface typically enhances photonic interactions and minimizes subsequent losses as photons solely propagate to and from the emitter. Emission enhancement into a selected, guided mode is quantified by the Purcell Factor $F$ \cite{purcell1995spontaneous}, which for PhCWs scales linearly with the group index $n_g$ \cite{hughes2004enhanced}. Consequently, PhCWs are often used in the slow-light regime \cite{schulz2010dispersion}, where $n_g \approx 58$ has been measured \cite{PhysRevLett.113.093603}. Unfortunately, in-plane backscattering between the counter-propagating modes scales as $n_g^2$ \cite{PhysRevLett.94.033903}, resulting in prohibitively large losses at high $n_g$'s. As we discuss below, the backscattering loss also depends on the way in which the electric field is distributed within the PhCW unit cell, opening up a route towards realistic slow-light interfaces.

In this work, we study how well four different PhCWs perform in each of the three areas identified above: $D$, $F$, and the minimization of backscattering losses. The corresponding unit cell of each structure is shown in Fig.~\ref{fig:Schematic}b, with the interface (center) of each waveguide marked. These have been designed to guide light near $\lambda = \SI{930}{\nano \meter}$ (see Appendix~\ref{sec:PCGeometryDetails} for further design parameters) for use with high-quality self-assembled Indium Arsenide quantum dots in a Galium Arsenide membrane \cite{lodahl2015interfacing} with a fixed lattice constant $a=266~{\rm nm}$, but their design can readily be scaled~\cite{joannopoulos1997photonic} for use with any other quantum photonic platform. Similarly, we limit our designs to circular holes but note that in practice more complex and fabricationally-challenging shapes such as triangles \cite{PhysRevResearch.2.043109} or shamrocks \cite{sollner2015deterministic} are possible.

We consider two topological and two conventional PhCW designs.
The conventional waveguides are standard photonic crystal line-defect waveguide (W1) \cite{arcari2014near}, and a broken-symmetry GPW that has been optimized to work as a chiral interface \cite{mahmoodian2017engineering}. We compare these to topological PhCWs based on a photonic analog of the Quantum Valley-Hall effect (QVH) of two photonic topological insulators \cite{mehrabad2020chiral}, each characterized by a topological invariant, the Valley-Chern number, of opposite sign $C_{\nu} = \pm 1/2$~\cite{PhysRevB.101.054307}. In analogy with the electronic QVH insulators \cite{ezawa2008quantum}, we expect that the difference between these invariants (here 1) denotes the number of topological interface modes that span the bandgap, although recent experiments suggest the existence of spectral regions where the mode does not afford protection to sharp bends \cite{mehrabad2020chiral}. Regardless, we use QVH insulators as their guided modes are known to lie below the light lines and hence do not couple to the free-space continuum,  in contrast to topological PhCWs based on the photonic analog of the Quantum Spin-Hall effect, whose modes lie above the light line and are therefore leaky \cite{arora2021direct,PhysRevResearch.2.043109}. 
Here, we consider QVH waveguide designs formed by bearded-type interface (BIW) and zig-zag-type (ZIW) interfaces \cite{rechtsman2012observation} as shown in Fig.~\ref{fig:Schematic}b.

\section{\label{sec:Waveguides}Photonic Band Diagrams and Dispersion}
We begin by calculating the (photonic) band diagrams and corresponding electromagnetic field distributions for all four PhCWs, using commercially available finite element software (COMSOL Multiphysics; details of the numerical simulations can be found in Appendix \ref{sec:FEM}). 

\begin{figure}[H]
    \centering\includegraphics[ width=0.95 \linewidth]{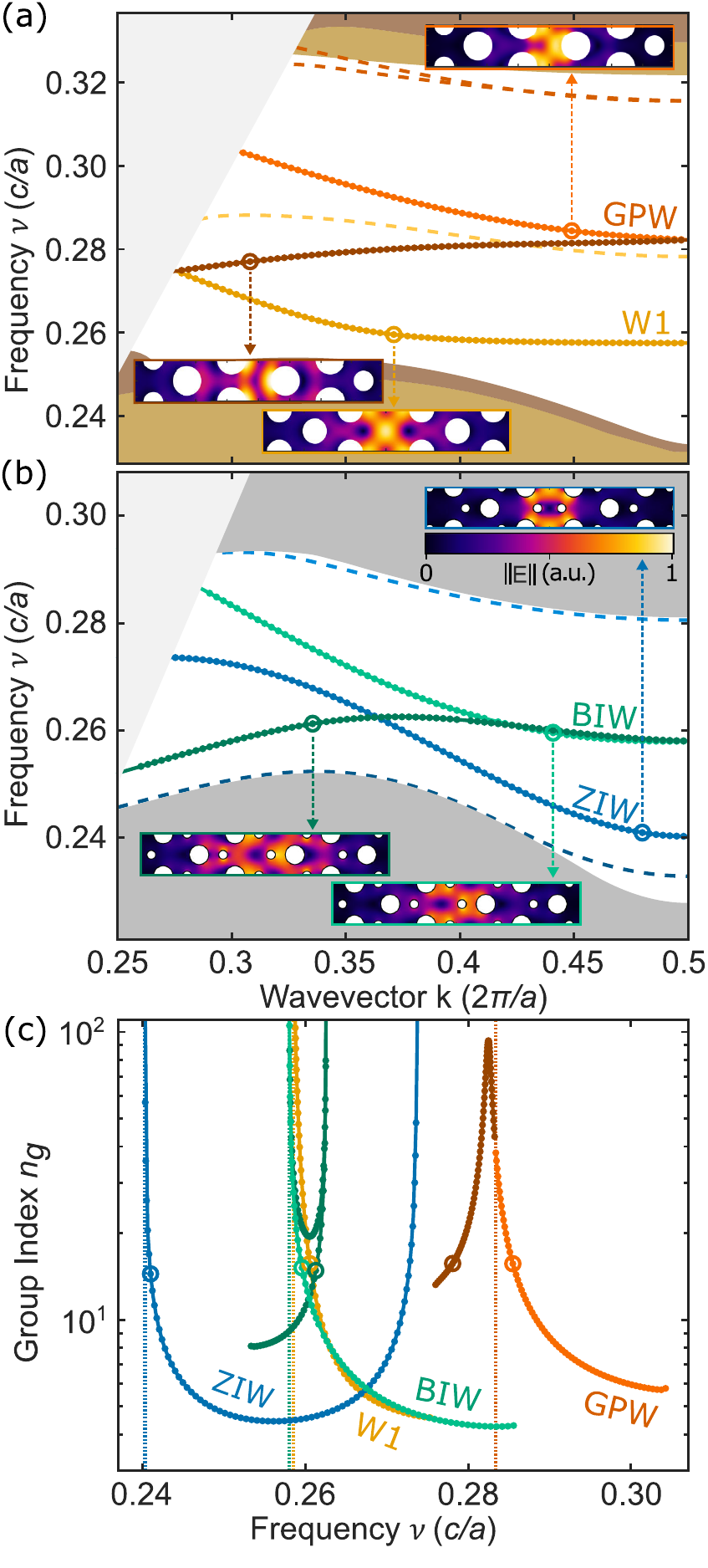}
    \caption{a) Photonic band diagram of the GPW (orange) and W1 (yellow) waveguides. Shaded regions correspond to the bulk modes of the GPW (brown) and W1 (yellow), while the grey region represents the light cone.
    The solid and dashed curves represent guided modes that are considered or excluded in this work, as discussed in the main text. Insets show exemplary mode profiles at $n_{g} \approx 15$. (b) Same as (a) but for topological BIW (green) and ZIW (blue) waveguides, noting that both topological waveguides share the same bulk modes. (c) Group index of the guided modes denoted by solid curves in (a) and (b) as a function of frequency for all four structures, with circles representing the modes whose profiles are shown above. The band-edges are indicated by a dotted line. 
    }
    \label{fig:BD_ng_all}
\end{figure}

We show the guided TE-like bands for the topological and conventional PhCWs 
in Fig.~\ref{fig:BD_ng_all}(a) and (b), respectively. In each case, the bulk continuum modes are given by the solid regions, while the guided modes  are given by solid curves. Additional modes that are close to the continuum and would therefore be leaky, or the higher-order mode of the W1 and GPW are shown by the dashed curves. We also show an exemplary Bloch normalized mode-profile of the electric field $\norm{\mathbf{e}_{n,k}}$ for each well-coupled mode of index $n$, taken for a group index  $n_{g}(\omega_{n,k}) = c / (\mathrm{d}\omega_{n,k} /\mathrm{d}k)  \approx 15$ [cf.~circles in Fig.~\ref{fig:BD_ng_all}(c)], where $c$ is the speed of light, $\omega_{n,k} = 2 \pi \nu_{n,k}$ is the eigenfrequecy, and $k$ is the wavenumber.
We note that dispersion engineering can both lift the entire guided bands of the topological photonic crystals into the bandgap~\cite{ma2016all, christiansen2019designing, christiansen2019topological} and avoid the mode-crossing of the BIW to recover single-mode operation across the entire $k$-space, as was done for the GPW~\cite{mahmoodian2017engineering}. In principle, further optimization of many of the PhCW's metrics is possible using inverse design \cite{nussbaum2021inverse}.

Several interesting similarities emerge between the dispersion relations of the topologically conventional and topological waveguides 
First, both the W1 and ZIW waveguides support one well-coupled mode and others that are poor choices for a quantum interface. The W1 dispersion contains an odd mode [dashed yellow curve in Fig.~\ref{fig:BD_ng_all}(a)] and the ZIW dispersion contains two modes in close proximity to the bulk modes [dashed blue curves in Fig.~\ref{fig:BD_ng_all}(b)] that in practice are expected to leak into the continuum. Furthermore, these two modes have a very large mode volume as discussed in the Appendix \ref{sec:FEM}, rendering them unsuitable for efficient light-matter coupling. The W1's fundamental mode is highly confined throughout the entire $k$-space, while the mode-volume of the ZIW mode has a more complex frequency dependence. As discussed in Appendix~\ref{sec:FEM}, the ZIW mode-width can be either large or small in regions of high $n_g$.


Similarly, there exist several similarities between the guided modes of the GPW and those of the BIW, as expected since both share the same broken transverse symmetry. The BIW topological edge-mode is comprised of two separate bands, each of which cover a large frequency interval, in much the same way as the guided modes of the GPW~\cite{mahmoodian2017engineering}. 
However, the two bands differ in their backscattering losses around bends as demonstrated in measurements and supported by finite-difference time-domain calculations \cite{mehrabad2020chiral}. For 60 degree bends, the upper mode shows little transmission,  while near-unity transmission was observed over a large bandwidth of the lower band, making it suitable for the creation of triangular resonators.
The two BIW bands cross at $ka/2 \pi \approx 0.42$ and are degenerate at the band edge. Both GPW and BIW modes are tightly confined across the entire $k$-space, with the mode-width of the BIW being significantly larger than that of the GPW and smaller than that of the ZIW, cf.~Appendix \ref{sec:FEM}. Nevertheless, we find highly confined, slow-light at the band edge of the lower branch of the BIW, indicating that large Purcell enhancement is possible with this structure.




\section{\label{sec:Merit} Directionality, Purcell Enhancement and Disorder-Induced Scattering}

\begin{figure*}[htb]
    \centering\includegraphics[width = 0.99\textwidth]{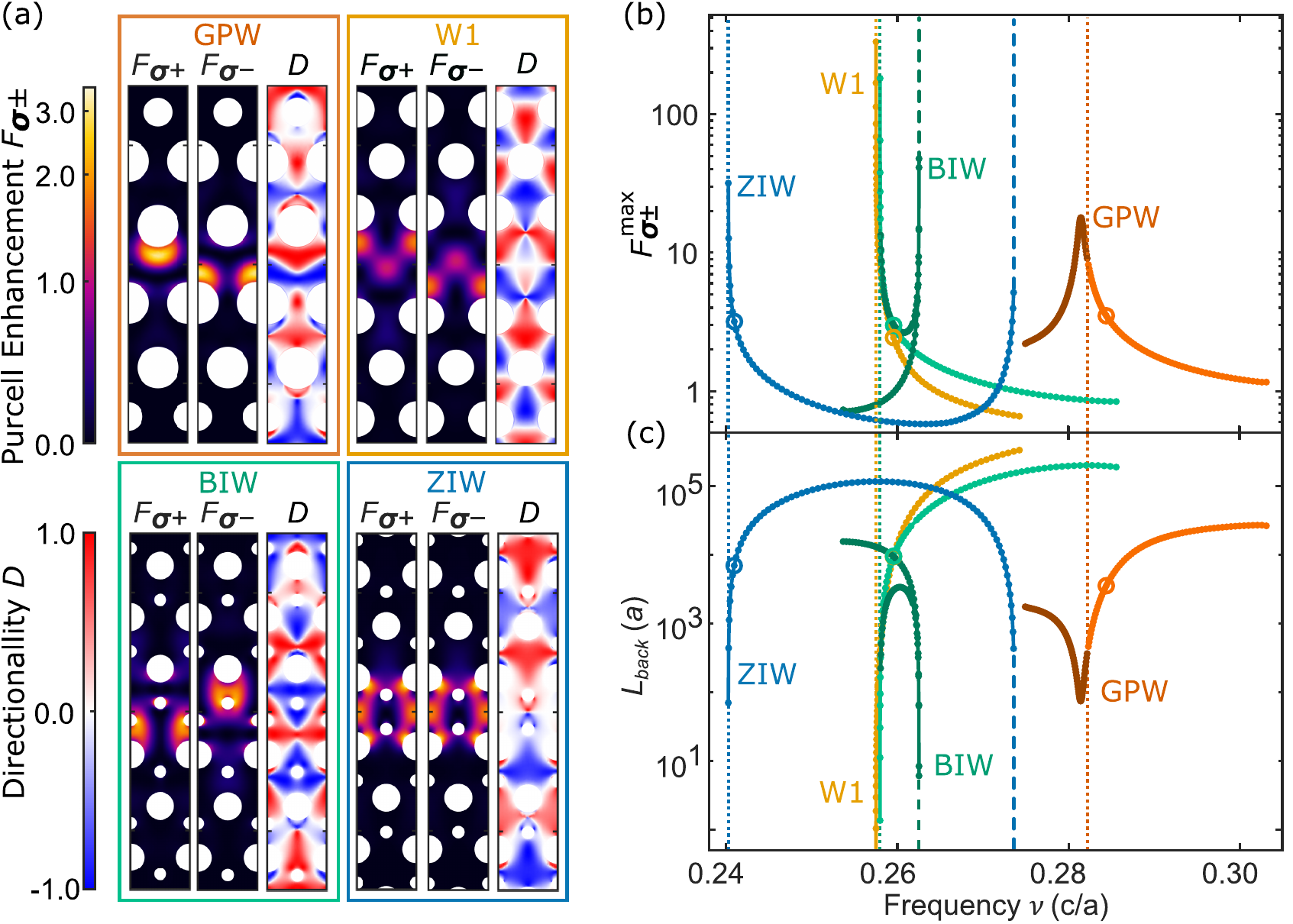}
    \caption{(a) Mode maps of the directional Purcell factor $F_{\boldsymbol\sigma_{\pm},k}(\mathbf{r})$  and directionality $D_{k}(\mathbf{r})$ for a unit cell of each PhCW, taken at $n_{g} \approx 15$ as shown in Fig.~\ref{fig:BD_ng_all}(c). (b) Maximal $F^{\rm max}_{\boldsymbol\sigma_{\pm},k}$ for each mode as a function of frequency, with dashed curves representing the divergence predicted in the slow-light regions. (c) Corresponding backscattering length ${L}_{{\rm back},n,k} $ in units of the lattice constant $a$ as a function of frequency. Note that both the backscattering losses and interaction enhancement diverge as the group index diverges. The band-edges are indicated by dotted lines, where the group index of all PhCWs expect for the GPW diverge. Dashed lines indicate group-index divergences of the band-edge. Circles in (b) and (c) represent the modes whose profiles are shown in (a).}
    \label{fig:ModePropeties}
\end{figure*}

Having determined the guided modes of each structures, we are now ready to quantify how well each functions as a bright, highly directional and low-loss interface. We begin by calculating the directional Purcell enhancement $F_{\boldsymbol\sigma_{\pm},n,k}$ for a left- (subscript $-$) or right-handed (subscript $+$) circular point-like dipole $\boldsymbol\sigma_\pm = 1/ \sqrt{2} \left(\hat{\mathbf{x}} \pm i \hat{\mathbf{y}} \right)$ for each PChW according to \cite{le2015nanophotonic} 
\begin{equation}
    F_{\boldsymbol\sigma_{\pm},n,k} \left( \mathbf{r} \right) =\frac{3 \pi c^2 a n_{g}\left( \omega_{n,k} \right) }{2 \omega_{n,k}^2 \sqrt{\epsilon \left( \mathbf{r} \right) }} \left| \boldsymbol\sigma_{\pm}^* \cdot \mathbf{e}_{n,k} \left( \mathbf{r} \right)  \right|^2.
\end{equation}
Although this quantity and others are mode dependent, since in what follows we only consider single-mode operation, the subscript $n$ is omitted for clarity. Exemplary maps of a unit cell of each structure are shown in Fig.~\ref{fig:ModePropeties}a, taken again for modes with $n_{g,n} \approx 15$. From these, we observe that for both the GPW and BIW there is little overlap between $F_{\boldsymbol\sigma_{+},k}$ and $F_{\boldsymbol\sigma_{-},k}$. In contrast, these maps differ only near the holes for the W1 structure, and are nearly identical for the ZIW, foreshadowing that these two structures fare poorly as chiral interfaces.


As noted above, the chirality of the interaction is quantified by the directionality,
\begin{equation}
    D_{k}\left( \mathbf{r} \right) =  \frac{F_{\boldsymbol\sigma_{-},k}\left( \mathbf{r}  \right) - F_{\boldsymbol\sigma_{+},k}\left( \mathbf{r} \right)}
    {F_{\boldsymbol\sigma_{+},k}\left( \mathbf{r}\right) + F_{\boldsymbol\sigma_{-},k}\left( \mathbf{r}  \right)},
\end{equation}
examples of which we also present in Fig.~\ref{fig:ModePropeties}(a). Here, we observe that while highly directional interactions are possible with all four structures, there is only a high degree of overlap between regions of high Purcell enhancement and directional interactions for the GPW and BIW. For the W1 and ZIW, in contrast, relatively high Purcell factors are found in regions of linear electric field polarization, largely precluding efficient chiral light-matter interactions and showing the importance of breaking inversion symmetry in the waveguides.


We find the maximal Purcell enhancement factor $F^{\rm max}_{\boldsymbol\sigma_{\pm},k} = \mathrm{max}_{\mathbf{r \in S_{c} }}\{F_{\boldsymbol\sigma_{\pm},k}(\mathbf{r})\}$ within each map (where $S_{c}$ denotes the area corresponding to the regions of high-index material), and plot these values as a function of frequency in Fig.~\ref{fig:ModePropeties}(b) for all structures.  We observe that all structures predict Purcell factors of upwards of 15 away from the band-edge, where both $n_g$ and $F_{\boldsymbol\sigma_{\pm},k}$ may diverge. In practice, the range of accessible factors is limited by several effects, among them scattering due to structural imperfections \cite{PhysRevB.72.161318} and disordered-induced mode broadening \cite{mann2015theory}).
Experimentally, a Purcell enhancement factor of $F \approx 6.9 $ has been observed \cite{PhysRevLett.113.093603}, and PhCW systems with significantly higher enhancements have been proposed \cite{rao2007single}.
From this perspective, all four structures, can theoretically enhance a chiral light-matter interaction sufficiently to form the basis of an efficient interface.

A good chiral interface must not only enhance interactions, but must also allow for subsequent low-loss transport. We therefore calculate the ensemble averaged
mean-free path (or backscatter loss length) ${L}_{{\rm back},k}  = \left\langle \alpha_{{\rm back},k}  \right\rangle^{-1}$, limiting ourselves to the single-mode and single-event backscattering regime as is typical for relatively short waveguides \cite{PhysRevB.80.195305}. Here, $\left\langle \alpha_{{\rm back},k}  \right\rangle$ is the  power-loss factor per unit cell that is the ensemble average over disorder-induced imperfections (see Appendix~\ref{sec:alpha} for details and discussion of the scattering mechanisms) for non-uniform air-hole size $R_\alpha$, where $\alpha$ is the index of the individual holes \cite{AdaptionTheory}. 
We neglect multi-mode scattering from degenerate modes of the BIW, since their degeneracy can be lifted without significantly altering their mode profile, as was done with the GPW \cite{mahmoodian2017engineering}, resulting in
\begin{align}
\left\langle \alpha_{{\rm back},k} \right\rangle & =   \sum_\alpha \frac{a^2 \omega_{k}^2 {n_{g}^2} \sigma^2}{4} (\epsilon_2 - \epsilon_1 )^2 \nonumber \\
& \times \iint \mathrm{d}\mathbf{r}  \mathrm{d}\mathbf{r}' \Theta\left(\frac{h}{2}-\abs{z}\right) \Theta\left(\frac{h}{2}-\abs{z'}\right) \nonumber \\
& \times \delta\left(R_\alpha - \left| \boldsymbol\rho - \boldsymbol\rho_\alpha \right| \right) \nonumber \\
&  \times  \left[ \mathbf{e}_{k}^{*} (\mathbf{r}) \cdot \mathbf{p}_{k}^{*} (\mathbf{r}) \right] \left[ \mathbf{e}_{k} (\mathbf{r}') \cdot \mathbf{p}_{k} (\mathbf{r}') \right] \nonumber \\ 
&  \times \mathrm{exp} \left( \frac{-R_\alpha \left| \Tilde{\phi} - \Tilde{\phi}' \right|}{l_p} + i2k(x-x') \right),
\end{align}
where $\sigma$ is the statistical surface roughness factor, $\Theta$ denotes the Heaviside function, $h$ is the membrane height, $z$ is the vertical Cartesian coordinate, $\delta$ is the Kronecker delta function, $\mathbf{\rho}$ ($\mathbf{\rho}_\alpha$) denotes the in-plane projection of the position vector (the individual hole center axis);
also, $\mathbf{p}_{k}$ is the polarizability, $l_p$ is the surface roughness correlation length, and $\Tilde{\phi}$ is the azimuth angle of the position vector in the cylindrical coordinate system which is centered in the hole $\alpha$. In principle, multi-mode back-scattering can be explicitly included \cite{PhysRevB.72.161318}. Yet is not expected to significantly contribute to losses other then for slow, small $k$ (leaky) modes \cite{parini2008time,PhysRevB.72.161318}.

\begin{figure*}[ht!]
    \centering\includegraphics[scale=0.99]{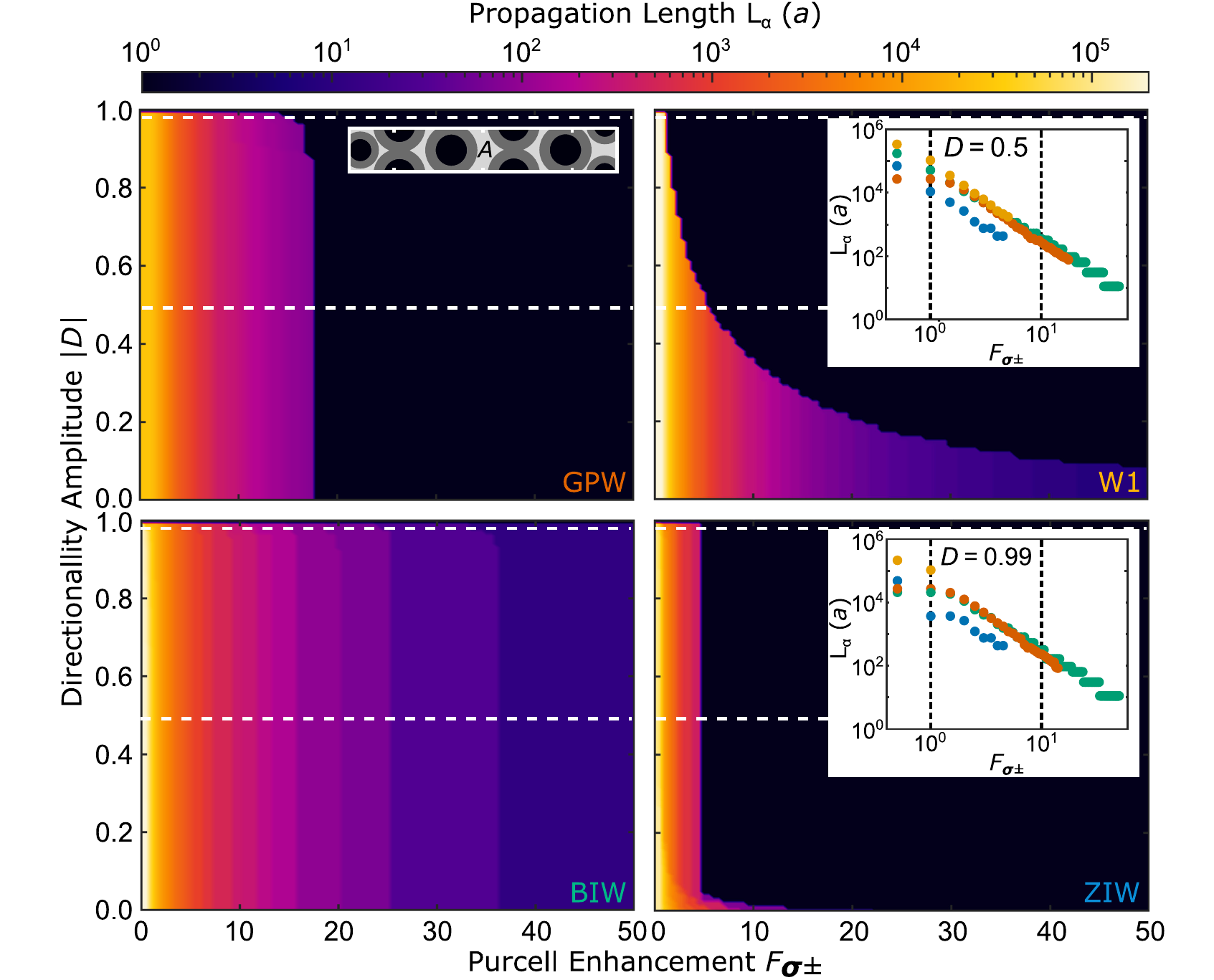}
    \caption{A summary of PhCW paramaters for use as practical quantum chiral-light matter interfaces. Maps of the maximal achievable propagation length (in units of the lattice constant $a$) as a function of the minimum desired directionality or Purcell Enhancement factor (structure marked in each panel). This combination of $D$ and $F$ must be found in an area $A \geq A_\mathrm{min}$, which is sufficiently distant $\delta_{min}$ from a hole, as sketched in the inset of the first panel and explained in the main text. The bright grey area indicates the area $A$ as an example for the W1, while the dark and black areas indicate the excluded regions given by $\delta_{\rm min}$. Insets in the W1 and ZIW panels show line cuts showing the propagation length as a function of $F$ for $\abs{D}=0.5$ and $\abs{D}=0.99$, respectively (cuts taken along the white dashed lines).}
    \label{fig:MinAlphMinFaMaps}
\end{figure*}

Considering state-of-the-art soft mask nanofabrication methods~ \cite{skorobogatiy2005statistical,PhysRevB.80.195305} we assume a surface roughness for each hole of $\sigma = \SI{3}{\nano \meter}$ and a correlation length for this disorder within each hole of $l_p = \SI{40}{\nano \meter}$ (cf.~Fig.~\ref{fig:lp_sweep}(b) and present the resulting mean-free path  ${L}_{{\rm back},k} = \langle \alpha_{{\rm back},k} \rangle $ for all structures in Fig.~\ref{fig:ModePropeties}(c). Interestingly, and as we show in Appendix~\ref{sec:alpha}, the relative performance of the different structures, with respect to backscattering losses is relatively insensitive to the absolute value of $l_p$. As expected, scattering losses increase with the group index, yet the absolute scattering length can significantly differ for the different structures (and in general does not scale quadratically with $n_g$~\cite{PhysRevB.80.195305}, as we also show in Appendix~\ref{sec:alpha_ng}). As an example, for $n_g = 15$, we find the Purcell factors and mean-free paths as listed in table \ref{tbl:example_L_F}. 
\begin{table}
\centering
\begin{tabular}{ c | c | c }
 PhCW & $F_{\sigma_\pm}^{\mathrm{max}}$  & ${L}_{\rm back}$ \\ 
 \hline
 W1 & $2.3$ & $ 2.4 \times 10^{6}~a$ \\  
 GPW & $3.4$ & $0.9 \times 10^{6}~a$  \\  
 ZIW & $1.6$ & $ 1.5 \times 10^{6}~a$  \\  
 BIW & $2.3$ & $2.2 \times 10^{6}~a$     
\end{tabular}
 \caption{Purcell factors $F_{\sigma_\pm}^{\mathrm{max}}$ and mean-free paths ${L}_{\rm back}$ for the highlighted modes with a group index of $n_g = 15$.}
 \label{tbl:example_L_F}
\end{table} 
That is, for this moderate group index all PhCWs show Purcell factors and mean-free paths varying by more than a factor of $2$. The W1 waveguide shows the least losses, while the GPW provides the strongest enhancement factor. This is to be expected, as can be seen in the mode distributions (cf.~Fig.~\ref{fig:BD_ng_all}(a), the field of the W1 PhCW is mainly located away from the holes (see Appendix \ref{sec:alpha_ng}), yet is only weakly circularly polarized (Fig.~\ref{fig:ModePropeties}(a). In contrast, while the GPW mode is more strongly localized near the hole edges, leading to higher scattering losses, yet is also circularly polarized at these areas of high field intensity.

While the BIW's performance according to table \ref{tbl:example_L_F}  seems similar to the  W1, the mode profile shown in Fig.~\ref{fig:ModePropeties} is very alike the GPW's. The lower Purcell factors and the higher losses stem from a wider Bloch mode and (See Appendix \ref{sec:FEM}) the high field strength at the hole-interfaces. Analogously, the ZIW's mode profile is similar to the W1's, but its wider Bloch mode results in the same performance reduction.

\section{\label{sec:Benchmark}Overall performance of the chiral interfaces }
A real quantum chiral light-matter interface must not only simultaneously enhance the interactions with circular transition dipoles and limit subsequent transport losses, but also be designed such that high-quality quantum emitters can be readily embedded in regions where coupling is effective.  This places two constraints: (i) the emitters cannot be located too close to the air-dielectric interfaces of the holes so as to avoid interactions with surface states, and (ii) the area of the region where the emitters can be located should be large enough to ensure a high yield of successful couplings (for details on the spatial constrains see Appendix \ref{sec:MinAlphMinFaMaps}). Condition (i) can typically be met with a distance $\delta_{\mathrm{min}} = 40~\mathrm{nm}$ \cite{gmeiner2016spectroscopy,liunanoscale,PhysRevLett.114.017601}, while state-of-the-art nanofabrication protocols allow for the deterministic solid-state emitter-photonic structure integration with an accuracy of around $\delta_{\mathrm{acc}} \approx 40~\mathrm{nm}$  \cite{schnauber2018deterministic,gschrey2015resolution,pregnolato2020deterministic}.

In Fig.~\ref{fig:MinAlphMinFaMaps} we therefore show the maximal propagation length $L_{\rm back}^{\rm max} \left( F_{\boldsymbol\sigma_{\pm}}, \abs{D} \right)$ that is possible, for a minimum desired diretionality amplitude $\abs{D}$ \cite{DirectionallityTRS} and Purcell enhancment $F_{\boldsymbol\sigma_{\pm}}$ for each structure with free choice of the mode and wavenumber $k$ \cite{ScaleInvariant}, but with the condition that a quantum emitter fits within an area  $A_{k}(F_{\boldsymbol\sigma_{\pm}},\abs{D},\delta_{\rm min})$ of minimum size $A_{\rm min} = \pi \delta_{\mathrm{acc}}^2$ while all point within this area are at least
$\delta_{\mathrm{\rm min}}$ away from the edge of an air hole (for details on the numerical implementation see Appendix \ref{sec:MinAlphMinFaMaps}). From this figure it is evident that only the GPW and BIW will realistically make good chiral light-matter interfaces, as there essentially does not exist a sufficiently large enough area to couple an emitter to either a W1 or ZIW structure with high directionality and even a moderate  $F_{\boldsymbol\sigma_{\pm}} =5$ ($F_{\boldsymbol\sigma_{\pm}} =3$) for the ZIW (W1) (although, for a ZIW, these points do exist within $\delta_{\mathrm{min}}$ of the air holes). In contrast, sufficiently large areas can be found within unit cells of the GPW and BIW where both near-perfect directionality $\left(\abs{D} \geq 0.99\right)$ and enhancements up to $F = 14$ and beyond are possible.



%

Where the topological BIW distinguishes itself from the GPW is both in the size of the area that can be used to efficiently and chirally interface to emitters, and in its performance at enhancement factors exceeding $14$.  For fixed $\abs{D}=0.99$ and $F_{\boldsymbol\sigma_{\pm}} \leq 10$ areas of at least $7.6\times10^5$~nm$^2$  (corresponding to a circle of radius $R_{\abs{D}=0.99}  \approx 491$~nm) can be found within a GPW unit cell, while for this area a BIW is limited to $4.4\times10^4$~nm$^2$ (corresponding to a circle of radius $R_{\abs{D}=0.99} \approx 118$~nm). This can be seen 
in Appendix~\ref{sec:MinAlphMinFaMaps}, where we also discuss how this area is determined.

Figure~\ref{fig:MinAlphMinFaMaps} shows that for high directionallity $\abs{D}=0.99$ the GPW outperforms the BIW as a chiral interface by about $30~\%$ higher propagation lengths, if low interaction enhancements $F_{\boldsymbol\sigma_{\pm}} \leq 4$ are desired. 
For intermediate Purcell enhancements ($4 \leq F_{\boldsymbol\sigma_{\pm}} \leq 12$) the GPW and the BIW show similar losses, while for higher interaction enhancements ($F_{\boldsymbol\sigma_{\pm}} > 12$) the propagation length of the BIW is longer than that of the GPW.  The highest achievable Purcell enhancement for the GPW is $F_{\boldsymbol\sigma_{\pm}}=17.5$ ($14$) for a minimal directionality amplitude $\abs{D}=0.5$ ($\abs{D}=0.99$). It is important to recognize that this limit arises due to the dispersion engineering of the GPW, which causes its group index to remain finite in the entire reciprocal space, in contrast to an unoptimized GPW whose dispersion relation diverges at the lower band edge~\cite{mahmoodian2017engineering}.

The difference between the losses of the GPW and BIW can be understood by considering their respective mode profiles (c.f. Appendix \ref{sec:FEM}), and realizing that in general the light is better confined in the GPW compared to the BIW. Particularly for low to intermediate group indices we find that the mode width of the GPW is half, or less, than that of the BIW 
(cf.~Appendix \ref{sec:FEM}). In the GPW, we can thus find the same enhancement as in the BIW for lower $n_g$'s, and hence less scattering losses. However, for large group indices the Bloch mode profile of the GPW's shows high intensity near the first air-hole row leading to a strong relative increase of the backscattering losses (c.f. Appendix \ref{sec:alpha_ng}). In other words, for low Purcell enhancements the GPW's disadvantage of high backscattering losses are compensated by tighter mode confinement. For example, for $\abs{D}=0.99$ we find a propagation length of $20880a$ ($51330a$) for a Purcell enhancement of $F=1$, and $318a$ ($240a$) for a Purcell enhancement of $F=10$ for the BIW (GPW).

For a minimal directionality amplitude of $\abs{D}=0.99$, we find a propagation length of $20880a$ ($51330a$) for a Purcell enhancement of $F=1$ and $318a$ ($240a$) for a Purcell enhancement of $F=10$ for the BIW (GPW). This means less than $\SI{4}{\percent}$ ($\SI{5}{\percent}$) backscattering losses for a $10$ unit cell long waveguide and out-of-plane scattering dominating for low Purcell enhancements \cite{patterson2009disorder}.

\section{Conclusions}
\label{sec:Conclsion}
From the selected 
photonic crystal waveguides studied,  only the GPW and the topological BIW are suitable platforms for chiral quantum optics. Both of these designs can enhance highly directional interactions for emitters located in relatively large regions. Both designs suffer from backscattering due to fabrication impurities, although the topological waveguide offers protection to 60-degree bends \cite{PhysRevLett.126.027403,yang2018topological,yamaguchi2019gaas,shalaev2019robust}, enabling sharp-edges microresonators that do not suffer from bending losses \cite{ma2016all,jalali2020semiconductor,mehrabad2020chiral,barik2020chiral,rechtsman2013photonic}. 


The ability of these structures to act as elements in viable quantum chiral light-matter interfaces is perhaps easier to visualize using a concrete example. Here, we consider the performance of a single quantum emitters chirally interfaced with a PhCW. This fundamental element enables realization of integrated non-reciprocal single-photon devices for constructing of scalable complex quantum circuits and networks \cite{lodahl2015interfacing}. Examples are loss-tolerant two-qubit measurements \cite{mahmoodian2016quantum}, optical isolators and circulators \cite{xia2014reversible,sollner2015deterministic}, photon number dependent routing \cite{chang2014quantum,volz2012ultrafast,javadi2015single}, $\sqrt{\rm{SWAP}}$ \cite{young2015polarization} and $\rm{CNOT}$ gates \cite{koshino2010deterministic}.


Specifically, we asses the setting where single photons are injected into the PhCW in which the chirally coupled quantum emitter coherently (and asymmetrically) scatters the injected photon forward or backwards \cite{lodahl2017chiral}. The forward propagating photons are single-sidedly collected after transmission through the PhCW of $N_\alpha$ unit cells.

The speed with which such a element ideally operates is set by the characteristic rate of the emitters, namely, their decay rates: $\gamma_{R/L} = \gamma_{R/L}^0 F_{\mathbf{\sigma},\pm}$, where $\gamma_{R/L}^0$ is the quantum emitter's directional emission rate in a homogeneous medium. In reality, the circuit operation rate (and fidelity) will be further decreased as photons are lost to scattering due to imperfections or imperfect directionality. Assuming backscattering only, the intensity, or accessible photon flux $\Phi_{R/L}$,  is thus given by,
\begin{equation}
    \Phi_{R/L} = \gamma_{R/L}^{0} F_{\boldsymbol\sigma_{\pm}} \mathrm{exp}\left(  - \frac{ N_a } {{L}_{\mathrm{\rm{back}}}^{\mathrm{max}} \left(F_{\boldsymbol\sigma_{\pm}},\abs{D}\right)} \right),
\end{equation} 
where we explicitly note that the maximum propagation length (in units of $a$) is a function of the desired enhancement and directionality.



\begin{figure}[t!]
    \centering\includegraphics[scale=0.98]{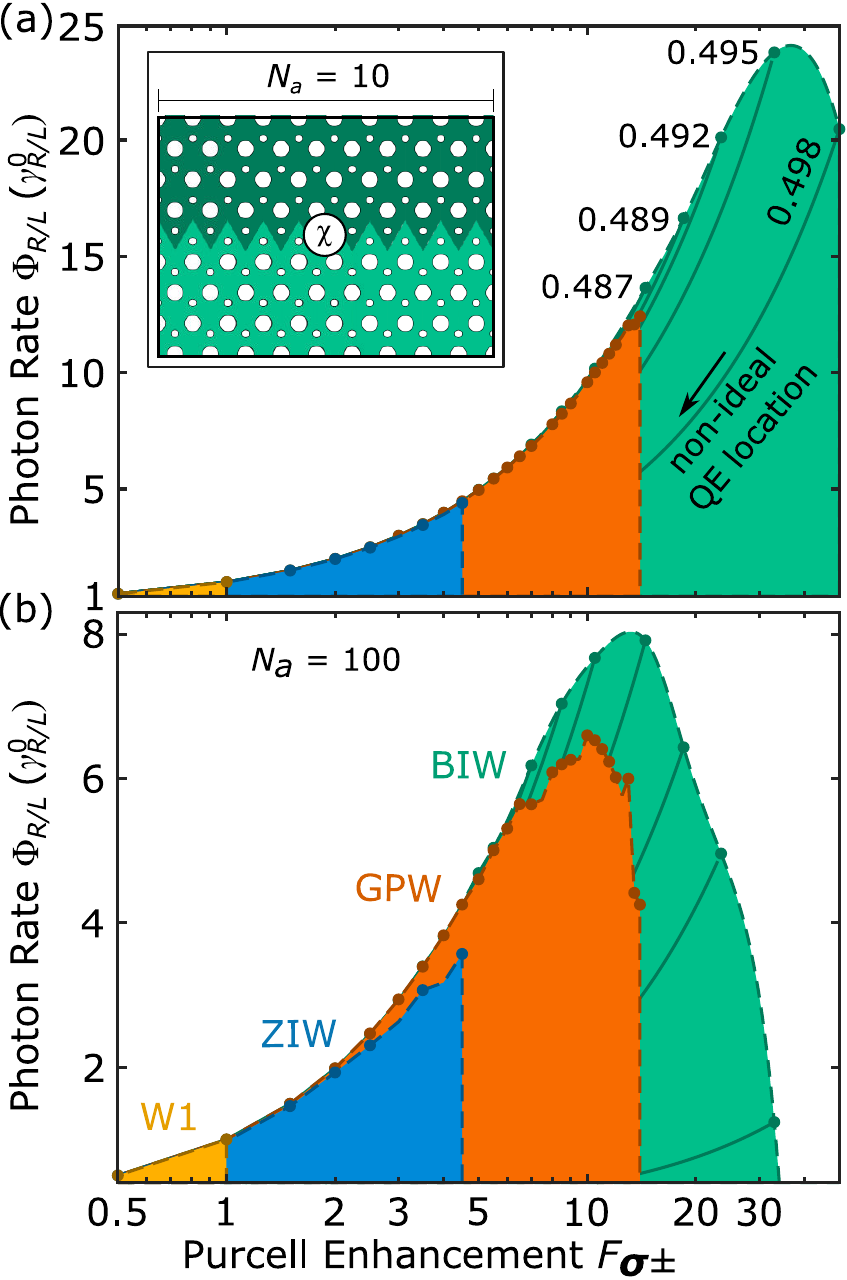}
    \caption{Calculated delectable photon flux, in units of the homogeneous decay rate, as a function of the Purcell enhancement for a quantum emitter $\chi$ chirally coupled to a PhCW as shown in the inset to (a) \cite{mahmoodian2016quantum}. The photon rate is shown for (a) short, 10 unit cell structures and (b), the more typical 100 unit cell waveguides in solid circles resolution limited by the $k$-space sampling, which are labeled as an example for the BIW for the data points of highest Purcell enhancements, c.f. Appendix \ref{sec:BandDiagramsForPhotonNumberRate}). The dashed lines represent a guide to the eye, while the solid curves represent a sub-optimal positioning of the QE. In both (a) and (b), the performance of all 4 structures is shown, with the topological BIW supporting highest-rate operation. This is true both for the shorter systems, where losses are low in all cases, as well as for the longer structures, where the high $n_g$ (enhancement) operation is limited by the scattering.}
    \label{fig:CountRate}
\end{figure}

We set $\abs{D}=0.99$ as for near ideal chiral coupling. We consider two limiting cases, a 10 unit cell PhCW ($\approx 2.7~\mu$m for our structures)  which is the shortest length to act as a proper interface \cite{rao2007single} and a 100 unit cell waveguide ($\approx27~\mu$m) which is a more typical length in current circuits \cite{sollner2015deterministic}, and plot $\Phi_{R/L} / \gamma_{R/L}^{0}$ as a function of $F_{\boldsymbol\sigma_{\pm}}$ in Fig.~\ref{fig:CountRate}a and b, respectively. The performance of the shorter structures is limited by the maximum achievable Purcell Factor (c.f. Fig.~\ref{fig:MinAlphMinFaMaps}), and we observe losses of less than $1\%$ for both GPW and BIW for $F \leq 2$ which increase to $10\%$ at around $F=13$. For higher values of $F$ operation is only possible with the BIW, and we expect $15\%$ losses at $F=20$. That is, for the shorter interfaces, low loss operation is compatible with moderate interaction enhancement. Here, we note that  $\Phi_{R/L}$ can directly quantify device performance, with the exact dependence determined by how many chiral elements are needed and the specific protocol (e.g. whether it depends on emission or transmission). For example, the operation speeds of quantum networks for all optical routing of single photons \cite{shomroni2014all} and loss-tolerant two-qubit measurements capable of universal quantum computation  \cite{mahmoodian2016quantum} is determined by the accessible photon flux directly, $\Phi_{R/L}$,  showing that GHz rate operation is possible.

For the longer devices we calculate about $1\%$  losses for the BIW and GPW, respectively, at $F=2$, which increase to $27\%$ and $34\%$ at $F=10$. At $F=20$, where only the BIW can be used, we expect upwards of $79\%$ losses, resulting in characteristic photon rates of $ \Phi_{R/L} / \gamma^0_{R/L} \approx 4$. Thus while even the longer waveguides can serve as chiral elements in quantum photonic circuits, for most efficient operation one should opt for a short, topologically protected waveguide.

However, for optimal operation as chiral light-matter interface the BIW requires further photonic engineering. Single-mode operation can be achieved analogous to the band engineering of the GPW \cite{mahmoodian2017engineering,he2019silicon}. Here, the BIW's additional degree of freedom allows can be exploited to simultaneously reduce backscattering losses while maintaining the access to modes of slow light and high directionality.

\begin{acknowledgments}
We thank Leonardo Midolo, for valuable discussions.
This project has received funding from the European Union's Horizon 2020 research and innovation programmes under grant agreement No. 824140 (TOCHA, H2020-FETPROACT-01-2018) and No. 801199.
We acknowledge funding from the Danish National Research Foundation (Center of Excellence “Hy-Q,” grant number DNRF139).
We acknowledge funding from 
the Canadian Foundation for Innovation (CFI) and
the Natural Sciences and Engineering Research Council of Canada (NSERC).
\end{acknowledgments}

\appendix

\section{Photonic Crystals Parameters}
\label{sec:PCGeometryDetails}

\begin{figure}[ht!]
    \centering\includegraphics[ scale=0.99]{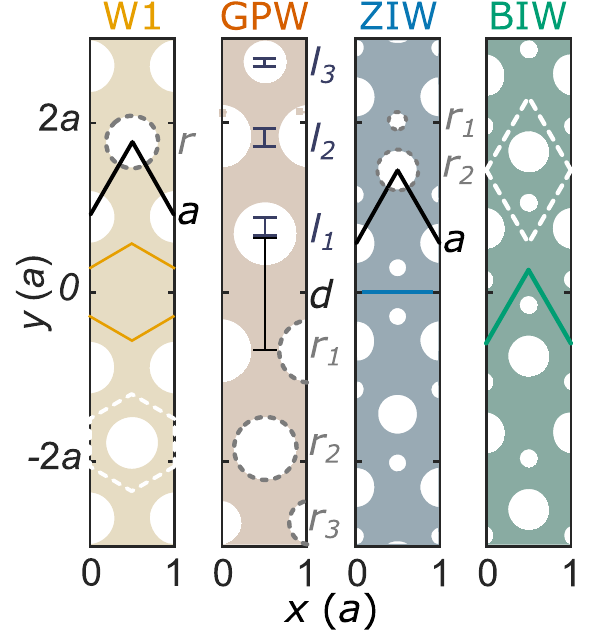}
    \caption{Schematic of the PhCWs's supercells and their parameters for the simulations, as discussed in the text. Each photonic crystal border lines are highlighted in color. Hole radii are indicate in grey. The GPW's photonic crystal is deformed from the W1 in the PhCW's center region as discussed in the text. The deformation parameters are indicated and labeld in grey. The lattice vectors (photonic crystal's unit cells) are indicated for the W1 and the topological waveguides in a solid black (white dashed) line. The ZIW's and BIW's photonic crystals have the same unit cell and lattice vectors but their interface follows along different directions.}
    \label{fig:Schematic_parameter}
\end{figure}

All PhCWs's parameters can be re-scaled with the crystal lattice constant $a$. For numerical implementation we have set $a = \SI{266}{\nano \meter}$ and a membrane thickness of $h=\SI{170}{\nano \meter}$. We have set the refractive index of the high-index membrane to $n_{1} = 3.4638$ and in surrounded by vaccum $n_{2} = 1$, representing a GaAs-based platform in cryogenic conditions ($T = \SI{4}{\kelvin}$) \cite{adachi1985gaas}.

For the the W1 have set a hole radius of $r/a = 0.3$.\
For the GPW, we follow the proposed design parameters \cite{mahmoodian2017engineering}.  The center hole-to-hole distance is $d = 0.75\sqrt{3}~a$. The radii $r_i$ and outwards position shifts $l_i$ of the first four rows of index $i$ of air holes are $r_1/a = r_2/a = 0.35$, $r_3/a = 0.24$, $r_4/a = 0.3$, and $l_1/a = 0$, $ l_2/a = \sqrt{3} / 8$, $l_3/a = \sqrt{3}/10$, $l_4/a = \sqrt{3}/20$. The other rows are not shifted in position have a hole radius identical to a W1's hole radii of $r/a = 0.3$.\\

The topological photonic insulator's unit cell consist of two hole of radius of $r_1/a = 0.105 $ and $r_2/a = 0.235$.

\section{FEM calculations settings, bloch mode normalization, group index, and mode width calculations}
\label{sec:FEM}

\begin{figure}[htb]
    \centering
        \includegraphics[scale=0.99]{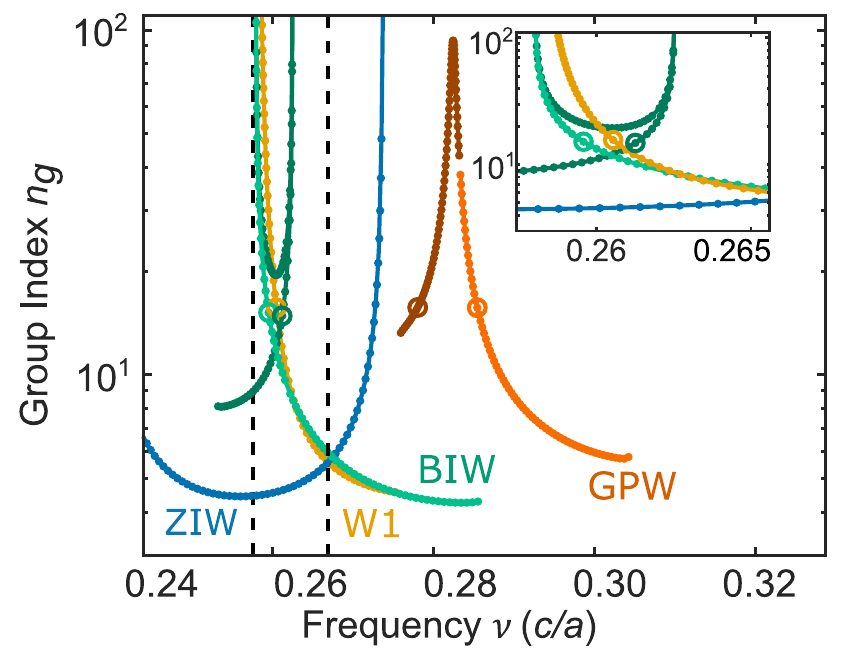}
    \caption{Group index $n_g(\omega)$ as function of the eigenfrequency of the selected modes of the BIW, ZIW, GPW, and W1 for comparison. The inset shows a detailed view of the frequency range indicated by dashed black lines. The circles indicate the mode profiles which are shown.}
    \label{fig:ng_detail}
\end{figure}

We performed numerical finite element calculations (FEMs) to determine the PhCW's Bloch modes and eigenvalues. The PhCWs are systems of mixed dielectric media for which we can compute the steady state solutions of the Maxwell equations resembling linear Hermitian eigenvalue problems \cite{joannopoulos1997photonic} by using the commercial software COMSOL Multiphysics. 

For each type of PhCW, a supercell of the one-dimensional periodic structure was studied using periodic boundary conditions (PBC). With a perfect magnetic conductor (PMC) plane, aligned with the reflection symmetry plane of the waveguides the numerical complexity is simplified limiting the study to transverse electric (TE) modes only. This takes the in-plane electric dipole moments of QEs suitable for chiral light-matter interaction \cite{lodahl2015interfacing} and TE-like mode profiles of the discussed PhCWs into account \cite{joannopoulos1997photonic,mahmoodian2017engineering,he2019silicon}. The simulated volume was restrained by perfectly matched layers (PML) below the membrane and each side of the PhCWs. The distance between the waveguide center and the PML below the membrane and on the sides of the PhCWs was chosen to achieve negligible mode leakage into the PML with a width of $12\sqrt{3}a$ on each side. Numerical convergence was tested sweeping the maximal element size of the mesh grid the Master's equation was solved for. This resulted in a tetrahedral mesh with a locally maximal element edge length of $1/10$ of the minimal distance between any local interface of differing dielectric constants. 


We defined a right-handed Cartesian coordinate system with $\hat{x}$ being aligned with direction of propagation in the waveguide, $\hat{y}$ pointininto the plane of the 2D photonic crystal $\hat{z}$ pointing out of plane, i.e., normal to the membrane.  

We sampled the $k$-space of the PhCW in increments of $\Delta k = 0.003  \cdot 2 \pi/a$ and computed the electric $\mathbf{e}_{n,k}(\mathbf{r})$ Bloch modes (with magnetic field   $\mathbf{h}_{n,k}(\mathbf{r})$) and eigenfrequncies $\nu_{n,k} = \omega_{n,k}(k) / 2 \pi$ of each mode Bloch mode $n$ of wavenumber $k$. For clarity we omit the mode index in the following. We also omit solutions in the light cone. We satisfy the Bloch mode normalization:
\begin{equation}
    \int_{\mathcal{V}_s} d\mathbf{r} \  \norm {\mathbf{e}_{k} \left( \mathbf{r} \right)}^2 \epsilon \left( \mathbf{r} \right) = 1,
\end{equation}
where \norm {\mathbf{e}_{k} \left( \mathbf{r} \right)} is the norm of the electric field of the Bloch mode and $\epsilon = n^2$ is the dielectric constant. We determine the group index for each mode $n_{g}(\omega_{k}) = c\frac{\mathrm{d}}{\mathrm{d}k} \omega_{k}$ using the Hellmann-Feynmann theorem \cite{joannopoulos2008molding}:
\begin{equation}
    n_{g}(\omega_{k}) = \frac{2c ( U_{\mathbf{e},k} + U_{\mathbf{h},k} ) }{\abs{ \int_{\mathcal{V}_s} d\mathbf{r} \ \mathrm{Re}(\mathbf{e}^*_{k}(\mathbf{r}) \times \mathbf{h}_{k}(\mathbf{r}))}},
\end{equation}
where $c$ is the speed of light, $\mathcal{V}_s$ is the total super cell volume, and $U_{\mathbf{e},k}$ ($U_{\mathbf{h},k}$) is the time averaged electric (magnetic) field energy.

We define and evaluate a characteristic mode width measure $w_{k}$ as the volume integral of the electric energy density $u_{\mathbf{e},k}(\mathbf{r})$ (magnetic $u_{\mathbf{h},n,k}(\mathbf{r} )$ density). We assigned the mode width measure $w_{k}$ implicit by the volume $\mathcal{V}_w = \left\{ \mathbf{r} \in \mathcal{V}_s  :-w_{k}<\mathbf{r} \times \hat{y}<w_{k} \right\}$ such that the time-averaged field energy inside the integration volume $\mathcal{V}_w$ is $1/e$ of the time-averaged field in the entire supercell volume $\mathcal{V}_s$:
\begin{equation}
    \int_{\mathcal{V}_w} \mathrm{d}\mathbf{r} \ \frac{u_{\mathbf{e},k}(\mathbf{r}) + u_{\mathbf{h},k}(\mathbf{r} )}{  U_{\mathbf{e},k} +U_{\mathbf{h},k}  } = \frac{1}{e},
    \end{equation}
where $u_{\mathbf{e},k}(\mathbf{r})$, $u_{\mathbf{h},k}(\mathbf{r} )$ are the electric and magnetic field energy density and $\mathcal{V}_w = \left\{ \mathbf{r} \in \mathcal{V}_s  :-w_{k}<\mathbf{r} \times \hat{y}<w_{k} \right\}$ is the effective mode volume. 

\begin{figure}[htb]
    \centering
        \includegraphics[scale=0.99]{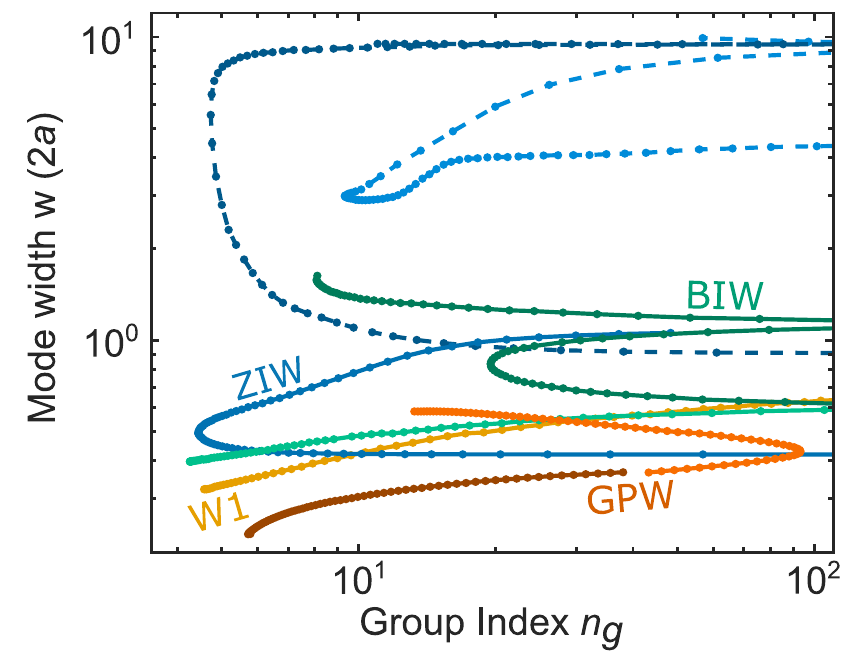}
    \caption{Characteristic mode width $w_{n}$ as function of the group index  $ {n}_{g}(\omega_{n,k})$ of all band-gap modes of the BIW, ZIW and the selected modes of the GPW and W1 for comparison. The data points are connected along their individual wavenumber $k$. The dashed lines correspond to the modes of the ZIW which are not considered for a chiral-light interface due to their extreme mode widths.}
    \label{fig:ModeWidthNg}
\end{figure}

The tight confinement of the GPW stems from the modified hole radii and position, and the modification of the space between the photonic crystals. The BIW's and ZIW's modes vary drastically in their characteristic mode width compared to topological trivial PhCWs and deviate clearly from the trend of increasing mode volume with increasing group index as observed for the W1. The GPW's lower frequency modes deviates from this trend only slightly.

\section{Incoherent  backscattering  power-loss  factor caused by structural disorder} 
\label{sec:alpha}

\begin{figure}[htb]
    \centering
    \includegraphics[scale=1.0]{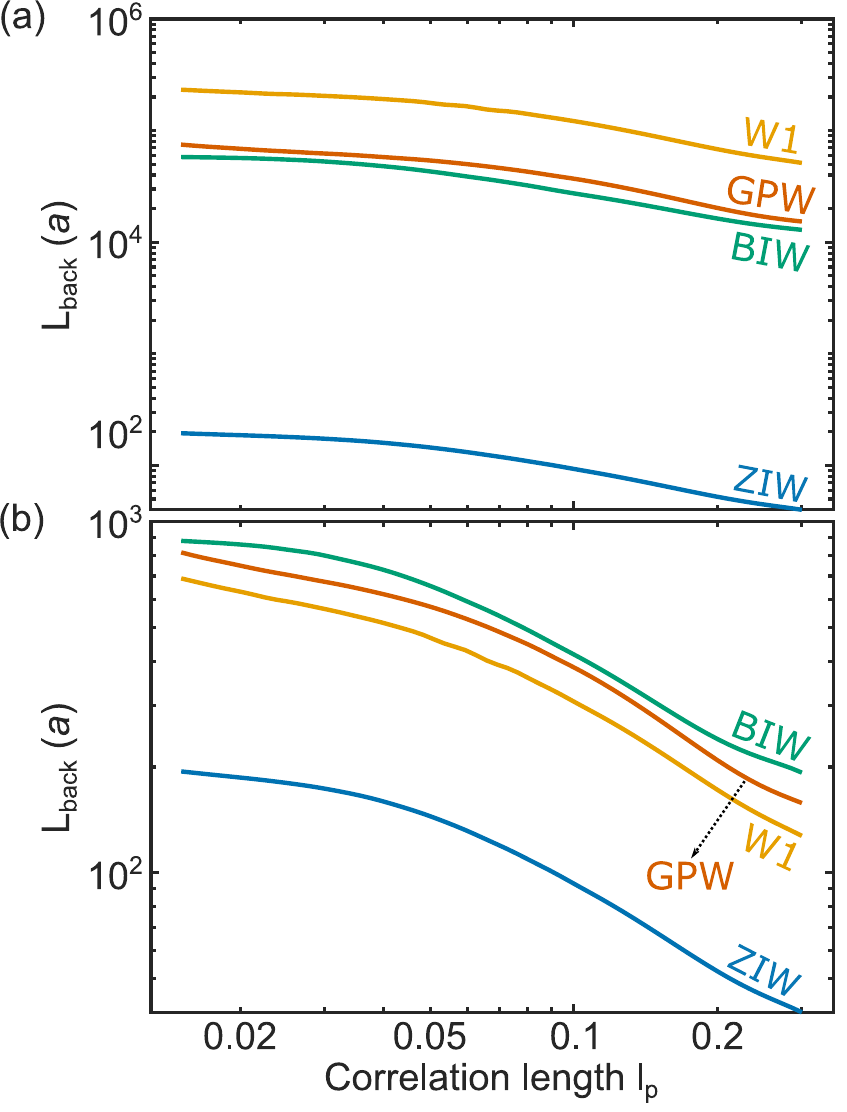}
    \caption{Inelastic mean-free path ${L}_{{\rm back},k}  = \left\langle \alpha_{{\rm back},k}  \right\rangle^{-1}$ sweeping the correlation constant $l_p$ and keeping the deformation parameter $\sigma=3~{\rm nm}$ constant for chosen modes  representing minimal backscattering fullfilling the requirements of: (a) providing Purcell enhancement $F=1$ and a directionality $\abs{D}=0.99$ in locations being separated from hole by $\delta$ within the high-index regions of the vertical symmetry plane $\mathcal{S}_c$ (see Appendix \ref{sec:MinAlphMinFaMaps}) and (b) providing a Purcell enhancement $F=10$ and providing the highest direcionality    $D_{\rm max}$ accessible for each PhCW within the same position restrictions as in (a).  The inlet shows a schematic of the stastical hole deformation parameters. A hole is deformed by a dent of amplitude $\sigma$ over a correlation length $l_p$.}
    \label{fig:lp_sweep}
\end{figure}

The ensemble average $\langle \alpha_{{\rm back}} \rangle$ of the fabrication disorder-induced single-event incoherent backscattering power loss per unit cell $\alpha_{{\rm back}}$ can be calculated semi-analytically using the slowly varying surface approximation \cite{PhysRevLett.94.033903, wang2008backscattering}. For a short W1 ($<\SI{100}{\micro \meter}$) and group index of about $n_g < 22$, $\alpha_{{\rm back}}$ is the backwards reflections loss described by the Beer-Lambert relation \cite{PhysRevB.72.161318}. For higher group indices or longer waveguides, the Beer-Lambert relation overestimates the backwards reflections for long waveguides due to multiple-scattering. In the multiple-scattering regime the effective losses per unit cell decreases to $\alpha_{{\rm back}}^{\textit{\rm eff}} = \alpha_{{\rm rad}} \sqrt{1+2\alpha_{\rm back}/\alpha_{{\rm rad}}}$ where $\alpha_{{\rm rad}}$ is radiative loss \cite{PhysRevB.72.161318}. For our discussion, we assume PhCWs of minimal length to form a chiral light-matter interface to achieve minimal losses. Therefore, we omit multiple-scattering events. PhCWs a short as $10$ unit cells are sufficient for high quality light-matter interfaces \cite{rao2007single}. We neglect the out-of-plane scattering losses, taking the scaling only linear in $n_g$ into account \cite{PhysRevB.72.161318}.

We also neglect multi-mode scattering as discussed in the main text and therefore find the backscattering power loss factor per unit cell by adaption from \citet{PhysRevB.81.245321} to be:
\begin{align}
\alpha_{{\rm back},k} & =  \frac{a^2 \omega_{k}^2 {n_{g}^2(\omega_{k})}}{4}   \nonumber \\ 
& \times \iint \mathrm{d}\mathbf{r}  \mathrm{d}\mathbf{r}' \Delta \epsilon \left( \mathbf{r} \right)  \Delta \epsilon( \mathbf{r}' )  \nonumber \\ 
& \times \left[ \mathbf{e}_{k}^{*} (\mathbf{r}) \cdot \mathbf{p}_{k}^{*} (\mathbf{r}) \right] \left[ \mathbf{e}_{k} (\mathbf{r}') \cdot \mathbf{p}_{k} (\mathbf{r}') \right]   \nonumber \\ 
& \times \mathrm{exp}\left(i2k(x-x')\right),
\end{align}
where $\Delta \epsilon( \mathbf{r})$ describes the difference of the dielectric function between the ideal and disordered structure and $\mathbf{p}_{n,k}$ is the polarization density:
\begin{equation}
    \mathbf{p}_{k}\left(\mathbf{r}\right) = \left( \mathbf{e}_{k,\parallel}(\mathbf{r}) + \epsilon(\mathbf{r})  \frac{\mathbf{d}_{k,\perp}(\mathbf{r})}{\epsilon_1 \epsilon_2}
    \right) \delta(\mathbf{r} - \mathbf{r}'),
 \end{equation}
where $\mathbf{e}_{k,\parallel}$ is the electric field components of the Bloch mode parallel to interfaces of changing dielectric constants $\epsilon_1 = n_1^2$ and $\epsilon_2 = n_2^2$ and $\mathbf{d}_{k,\perp}$ is perpendicular electric displacement fields. This disorder
form help to satisfy the correct boundary conditions at the hole interface.

We assume in-plane hole deformation of $\Delta R$ of the holes to be the dominant source of scattering in in good agreement with theory and experiment \cite{PhysRevB.72.161318, patterson2009disorder}. Thus $\Delta \epsilon( \mathbf{r})$ is only non-zero at the hole walls. With air hole indices $\alpha$ and the corresponding hole's Radius $R_\alpha$, we can write the change of the dielectric function as:
\begin{align}
    \Delta \epsilon \left( \mathbf{r} \right) & =  \left( \epsilon_2 - \epsilon_1 \right)  \Theta\left(\frac{h}{2}-\abs{z}\right) \sum_\alpha \Delta R \left( \Tilde{\phi}\left( \boldsymbol\rho,\boldsymbol\rho_\alpha \right) \right) \nonumber \\ & \times \delta\left(R_\alpha - \left| \boldsymbol\rho - \boldsymbol\rho_\alpha \right| \right),
\end{align}
where $\boldsymbol\rho$, $\boldsymbol\rho_\alpha$ are the in-plane vectors to $\mathbf{r}$ and the hole center position of hole $\alpha$, and where $\Tilde{\phi}$ is angular coordinate of the position $\mathbf{r}$ in the cylindrical coordinate system center in the hole $\alpha$, so that
\begin{equation}
    \Tilde{\phi} \left( \boldsymbol\rho,\boldsymbol\rho_\alpha \right) = \mathrm{arctan} \left( \frac{\boldsymbol\rho \mathrm{sin}\left(\phi \right) - \boldsymbol\rho_\alpha \mathrm{sin}\left(\phi_\alpha \right) }{\boldsymbol\rho \mathrm{cos}\left(\phi \right) - \boldsymbol\rho_\alpha \mathrm{cos}\left(\phi_\alpha \right)} \right).
\end{equation}

We assume disorder between different air holes $\alpha_i$ and $\alpha_j$ to be uncorrelated for $i \neq j$ but to perfectly correlated within each air hole in the cylindrical axis direction. This assumption takes the statistical functions determined by imaging of photonic crystal slabs into account \cite{skorobogatiy2005statistical}. We can write the the disorder correlation between two point of the sidewall as
\begin{align}
    \left\langle \Delta R \left( \Tilde{\phi} \right) \Delta R \left( \Tilde{\phi}' \right) \right\rangle & = \sigma^2 \mathrm{exp} \left( \frac{-R_\alpha \left| \Tilde{\phi} - \Tilde{\phi}' \right|}{l_p}\right) \nonumber \\
& \times \delta(\alpha,\alpha').
\end{align}

Thus, the ensemble averaged incoherent disorder-induced backwards scattering power loss per unit cell in the single-scattering event approximation omitting mutli-mode scattering is then: 
\begin{align}
\left\langle \alpha_{{\rm back},k} \right\rangle & =   \sum_\alpha \frac{a^2 \omega_{k}^2 {n_{g}^2(\omega_{k})} \sigma^2}{4} (\epsilon_2 - \epsilon_1 )^2 \nonumber \\
& \times \iint \mathrm{d}\mathbf{r}  \mathrm{d}\mathbf{r}' \Theta\left(\frac{h}{2}-\abs{z}\right) \Theta\left(\frac{h}{2}-\abs{z'}\right) \nonumber \\
& \times \delta\left(R_\alpha - \left| \boldsymbol\rho - \boldsymbol\rho_\alpha \right| \right) \nonumber \\
&  \times  \left[ \mathbf{e}_{k}^{*} (\mathbf{r}) \cdot \mathbf{p}_{k}^{*} (\mathbf{r}) \right] \left[ \mathbf{e}_{k} (\mathbf{r}') \cdot \mathbf{p}_{k} (\mathbf{r}') \right] \nonumber \\ 
&  \times \mathrm{exp} \left( \frac{-R_\alpha \left| \Tilde{\phi} - \Tilde{\phi}' \right|}{l_p} + i2k(x-x') \right).
\end{align}
This formalism recovers the approximate backscattering scaling quadratic in the group index. However, the backscattering highly dependence of the PhCW's morphology by means of the intensity profile at the air-hole walls, and the Bloch modes change as a function of frequency and $k$~\cite{PhysRevB.80.195305}. Mode profiles with high fields strengths at a large number of holes show high backscattering losses. Similarly, are smaller holes associated to larger backscattering losses due to the $R_\alpha / l_p$-term, as they occur in the third row of holes in the GPW and the throughout the BIW's and the ZIW's photonic crystals.

While all WG scale quadratric in $\sigma$ the effect of $l_p$ is non trivial, depending on the hole-sizes and the field amplitude at the air-hole interfaces. However, the influence of the backscattering parameters seems to be low and affects all PhCWs similarly as shown in figure \ref{fig:lp_sweep} where we show the  inelastic mean-free path ${L}_{{\rm back},k}  = \left\langle \alpha_{{\rm back},k}  \right\rangle^{-1}$. Thus, the choice of the correlation length and the deformation strength do not influence the choice of the waveguide topology significantly.
    
\section{Backscattering and Purcell enhancement scaling in the group index}
\label{sec:alpha_ng}   

Upon re-normalization of the Purcell enhancement and of the mean free path by their inverse explicit group index dependence, the significance of the dispersion of the mode profiles and their polarization is revealed. We observe variations in the mode dependence of the re-normalized enhancement (mean free path)
vary up to a factor $4$ ($20)$ throughout the full $k$-space outside the light cone. For identical group index the implicit group index dependence is taking into account. Still, we find observe variations in the mode profile dependence of the re-normalized enhancement (mean free path)
vary up to a factor $4$ ($9)$. The mode profile dependence of both properties does not correlate among different PhCW and their relative dependence is highly dispersive.

\begin{figure}[htb]
    \centering
    \includegraphics[scale=1.0]{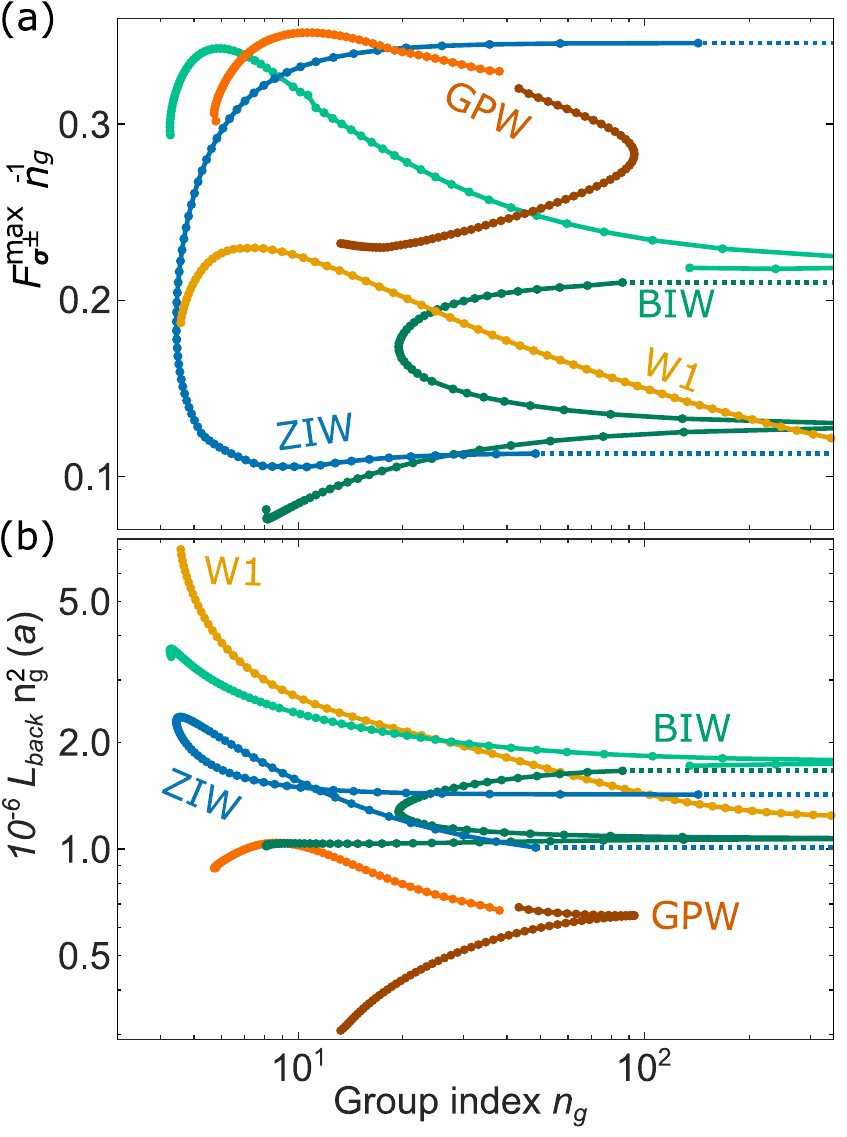}
    \caption{(a) Re-normalized maximal Purcell enhancement $F_{\boldsymbol\sigma_{\pm},k}^{\rm max} / n_g$  within the high-index regions of the vertical symmetry plane $\mathcal{S}_c$ (see Appendix \ref{sec:MinAlphMinFaMaps}) and (b) re-normalized mean free path ${L}_{{\rm back},k} n_g^2$. The solid lines connect modes along the wavenumber $k$ while the dashed lines indicate regions of divergence. The non-trivial dispersion of the re-normalized Purcell enhancement and the re-normalized mean free path indicate the significance of the dispersion of the mode profiles and their polarization.}
    \label{fig:alpha_ng}
\end{figure}


\section{Areas of Purcell enhancement and directionality}
\label{sec:MinAlphMinFaMaps}

\begin{figure*}[htb]
    \centering\includegraphics[width=0.99\textwidth]{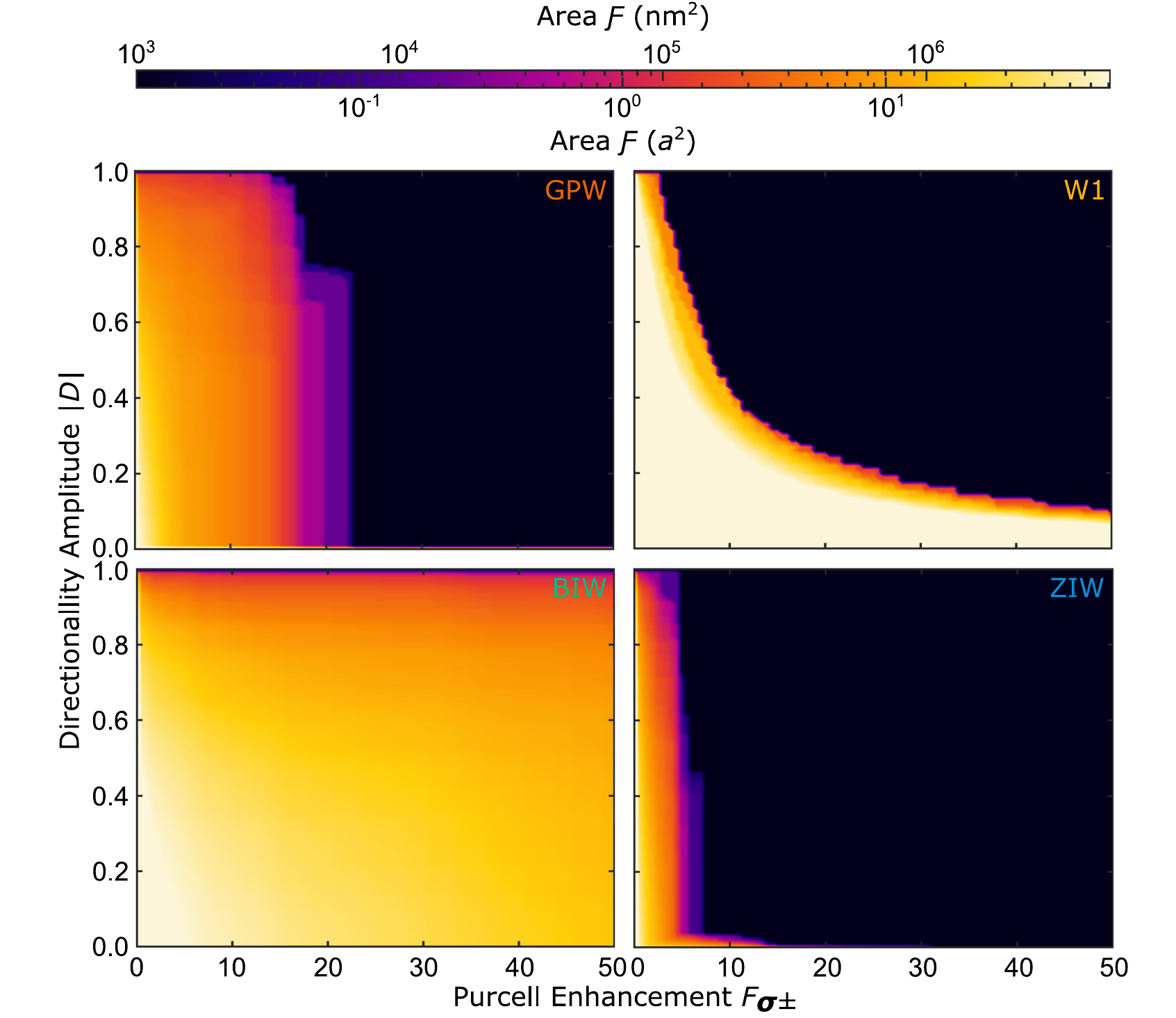}
    \caption{Maximal area $\mathcal{F}$ of all modes for minimal directionality amplitude $\abs{D}$ and minimal Purcell enhancements $P$ in the  high index material area given by $\mathcal{S}_{\rm min}$, not limiting the associated  group index or backscattering losses }
    \label{fig:MaxAreaMaps_d30}
\end{figure*}

In order to compare the PhCWs' performances as quantum emitter based chiral light-matter interface, we can compare the maximal propagation length ${L}_{{\rm back},n,k} $ under free choice of the mode and free choice of wavenumber $k$ for which we find a non-zero area in which we find a minimal directionality $D$ Purcell enhancement $F$. The free choice of wavector $k$ and mode $n$ means to tune the PhCW optimally to resonance with the QE by tuning the lattice parameters \cite{ScaleInvariant}. However, for a realistic platform, we have to take two more restrictions of QE positioning into account. State-of-the art fabrication techniques only allows for placing QE only at certain minimal distance $\delta_{\rm min}$ from any air-hole wall interface and only with a certain positioning accuracy $\delta_{\rm acc}$.

Experimentally, the minimum air hole distance has been estimated for near-infrared InGaAs QDs to be about $\delta_{\rm min} = \SI{30}{\nano \meter}$ \cite{pregnolato2020deterministic}. We take this restriction of positioning QDs mathematically into account by only considering points in the manifold $\mathcal{S}_{\rm min}$, defined as all points the symmetry plane of the high index material $\mathcal{S}_c$ that have a minimal distance $\delta_{\rm min}$ to any air hole:
\begin{equation}
   \mathcal{S}_{\rm min}(\delta_{\rm min}) = \left\{\mathbf{r} \in \mathcal{S}_c : \norm{\boldsymbol\rho - \boldsymbol\rho_\alpha } \geq R_\alpha + \delta_{\rm min} \right\},
\end{equation}
where $\mathcal{S}_c$ is the high index material region of the PhCWs vertical symmetry plane of the supercell, and $\boldsymbol\rho$ ($\boldsymbol\rho_\alpha$) is the in-plane vector to $\mathbf{r}$ (the center of hole with index $\alpha$). 

The area $A_{k}(\abs{D},F,\delta_{\rm min})$ of a minimal directionality $  D_{k}(\mathbf{r})  >\abs{D}$ Purcell enhancement $F_{\boldsymbol\sigma_{\pm},k} \left( \mathbf{r} \right) > F$ per supercell considering the QD positioning restriction given by $\delta_{\rm min}$ is:
\begin{align}
    A_{k}(\abs{D},F,\delta_{\rm min}) & = \int_{\mathcal{S}(\delta_{min})} \mathrm{d} \mathbf{r} \nonumber \\
& \times  \Theta\left(    F_{\boldsymbol\sigma_{+},k} \left( \mathbf{r} \right) - F \right)  \nonumber \\
& \times \Theta\left( D_{k}   \left( \mathbf{r} \right)  - \abs{D} \right),
\end{align}
where $\Theta$ is the Heaviside step function.\\

In figure \ref{fig:MaxAreaMaps_d30} we show the maximal areas $A_{k}(\abs{D},F,\delta_{\rm min})$ for a Purcell enhancement $F$ and Directionality $D$ under free choice of the mode and the wavenumber $k$.
\begin{equation}
    \mathcal{F}  = \mathrm{max}_{k}\{A_{k}(\abs{D},F,\delta_{\rm min})\}.
\end{equation}

\begin{figure}[ht]
    \centering
    \includegraphics[scale=1]{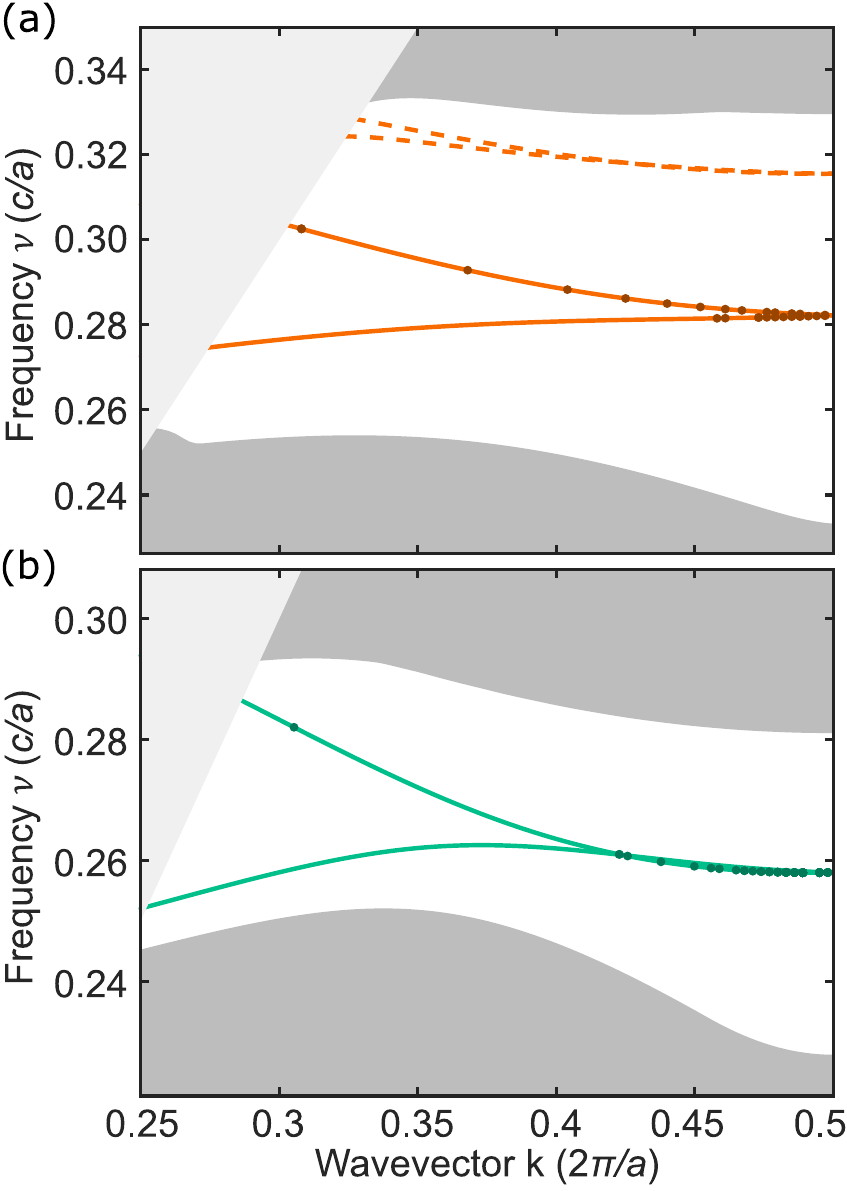}
    \caption{Band diagram of the (a) GPW and (b) BIW showing only the light cone, the bulk bands and the relevant modes for which the photon number rate $\Phi_{L/R}/\gamma_{L/R}^{0}$ is maximal, desiring a directionality of $\abs{D}=0.99$ and  minimal losses. Each point shown corresponds to a different optimal Purcell  enhancement.}
    \label{fig:ModesForMZI}
\end{figure}

We find the W1 to show large areas for high Purcell enhancements but due to its predominately linear polarization throughout the $k$-space, it does not allow for Purcell enhancements and high directionality for $F>3$. The ZIW is very similar to the W1 in this regard, although the areas $A$ are for all pairs $D,F$ at least a order of magnitude smaller. The GPW and BIW seem to similar. However, for areas as of $\abs{D} > 0.9$ ($\abs{D} < 0.9$) the GPW (BIW) offers more area for most pairs $D,F$. The BIW is the only PhCW allowing for QE interfaces with significant directionality and the highest Purcell enhancement of $F>20$.

We take the QE positioning accuracy $\delta_{\rm acc}$ into account by only considering areas $A_{k}(\abs{D},F,\delta_{\rm min})$  large enough to place a QD within the minimum area $A_{min}(\delta_{\rm acc}) = \pi \delta_{\rm acc}^2$. The precision of placing a InGaAs QD in a photonic structure is estimated to be about $\delta_{\rm acc} = \SI{40}{\nano \meter}$ \cite{pregnolato2020deterministic}.

Taking both fabrication limitations into account, we asses the maximal propagation length ${L}_{\rm back}^{\rm max}(F,\abs{D})$ under free choice of the wavenumber $k$ and free choice of the mode for which we find a area $A_{k}(\abs{D},F,\delta_{\rm min}) > A_{\rm min}(\delta_{\rm acc})$ of a minimal directionality amplitude $D_k(\mathbf{r}) > \abs{D}$ and minimal Purcell enhancement $ F_{\boldsymbol\sigma_{\pm},k} \left( \mathbf{r} \right) > F$ and take the QE positing limitations into account as:

\begin{align}
    {L}_{\rm back}^{\rm max}&(\abs{D},F,\delta_{\rm min},\delta_{\rm acc})  = \nonumber \\
    & \mathrm{max}_{k} \{  {L}_{{\rm back},k} : \nonumber \\ 
    & A_{k}(\abs{D},F,\delta_{\rm min}) > A_{\rm min}(\delta_{\rm acc}) \}.
\end{align}
We show the conditional maximal propagation length ${L}_{\rm back}^{\rm max}(\abs{D},F,\delta_{\rm min},\delta_{\rm acc})$ in figure \ref{fig:MinAlphMinFaMaps}.

\section{Modes for highest photon number rates}
\label{sec:BandDiagramsForPhotonNumberRate}

The modes for maximal photon number rate $\Phi_{L/R}/\gamma_{L/R}^{0}$ are for a directionallity $\abs{D}=0.99$ and minimal losses as function of the Purcell enhancement F are highlighted in figure \ref{fig:ModesForMZI}. The discreet sampling of the $k$-space allows only for discreet evaluation of the optimal mode with optical enhancement to loss-ratio. Particularly, for high enhancements involving high group indices, the finite sampling of the $k$-space limits the resolution of the optimal enhancement to loss ratio.

For increasing enhancement desired, the group index increases monotonously. When increasing the enhancements for the BIW the wavenumber $k$ monotonously increases as well. For the BIW only modes of the upper band are optimal. In case of the GPW the upper band shows less losses for the same group index compared to the lower band. However, the lower bands exceeds the upper bands maximal group index and thus its maximal accessible enhancement. Consequently, modes of the lower band with high losses are to be utilized above a certain desired Purcell enhancement.




\bibliographystyle{aipauth4-1}
\bibliography{ArXiv_Submission}

\begin{thebibliography}{88}%
\makeatletter
\providecommand \@ifxundefined [1]{%
 \@ifx{#1\undefined}
}%
\providecommand \@ifnum [1]{%
 \ifnum #1\expandafter \@firstoftwo
 \else \expandafter \@secondoftwo
 \fi
}%
\providecommand \@ifx [1]{%
 \ifx #1\expandafter \@firstoftwo
 \else \expandafter \@secondoftwo
 \fi
}%
\providecommand \natexlab [1]{#1}%
\providecommand \enquote  [1]{``#1''}%
\providecommand \bibnamefont  [1]{#1}%
\providecommand \bibfnamefont [1]{#1}%
\providecommand \citenamefont [1]{#1}%
\providecommand \href@noop [0]{\@secondoftwo}%
\providecommand \href [0]{\begingroup \@sanitize@url \@href}%
\providecommand \@href[1]{\@@startlink{#1}\@@href}%
\providecommand \@@href[1]{\endgroup#1\@@endlink}%
\providecommand \@sanitize@url [0]{\catcode `\\12\catcode `\$12\catcode
  `\&12\catcode `\#12\catcode `\^12\catcode `\_12\catcode `\%12\relax}%
\providecommand \@@startlink[1]{}%
\providecommand \@@endlink[0]{}%
\providecommand \url  [0]{\begingroup\@sanitize@url \@url }%
\providecommand \@url [1]{\endgroup\@href {#1}{\urlprefix }}%
\providecommand \urlprefix  [0]{URL }%
\providecommand \Eprint [0]{\href }%
\providecommand \doibase [0]{http://dx.doi.org/}%
\providecommand \selectlanguage [0]{\@gobble}%
\providecommand \bibinfo  [0]{\@secondoftwo}%
\providecommand \bibfield  [0]{\@secondoftwo}%
\providecommand \translation [1]{[#1]}%
\providecommand \BibitemOpen [0]{}%
\providecommand \bibitemStop [0]{}%
\providecommand \bibitemNoStop [0]{.\EOS\space}%
\providecommand \EOS [0]{\spacefactor3000\relax}%
\providecommand \BibitemShut  [1]{\csname bibitem#1\endcsname}%
\let\auto@bib@innerbib\@empty
\bibitem [{Ada()}]{AdaptionTheory}%
  \BibitemOpen
  \href@noop {} {\bibinfo  {journal} {Adapted from
  ref.~\citet{PhysRevB.81.245321} for non-uniform hole sizes $R_\alpha$}\
  }\BibitemShut {NoStop}%
\bibitem [{Dir()}]{DirectionallityTRS}%
  \BibitemOpen
\bibfield  {journal} {  }\href@noop {} {\bibinfo  {journal} {In all-dielectric
  media light propagation is described by Maxwell’s equations obeying
  time-reversal symmetry. In that regard the overall performance of the
  discussed chiral interfaces for a given diretionality $D$ is identical to a
  directionlity $-D$.}\ }\BibitemShut {NoStop}%
\bibitem [{Sca()}]{ScaleInvariant}%
  \BibitemOpen
\bibfield  {journal} {  }\href@noop {} {\bibinfo  {journal} {For photonic
  crystals there is no fundamental constant with the dimension of length since
  the master equation in dielectric media is scale invariant. A quantum emitter
  can be brought in resonance with a mode of wavenumber $k$ by scaling the
  lattice constant $a$}\ }\BibitemShut {NoStop}%
\bibitem [{\citenamefont {Adachi}(1985)}]{adachi1985gaas}%
  \BibitemOpen
\bibfield  {journal} {  }\bibfield  {author} {\bibinfo {author} {\bibnamefont
  {Adachi}, \bibfnamefont {S.}},\ }\href@noop {} {\bibfield  {journal}
  {\bibinfo  {journal} {Journal of Applied Physics}\ }\textbf {\bibinfo
  {volume} {58}},\ \bibinfo {pages} {R1} (\bibinfo {year} {1985})}\BibitemShut
  {NoStop}%
\bibitem [{\citenamefont {Arcari}\ \emph
  {et~al.}(2014{\natexlab{a}})\citenamefont {Arcari}, \citenamefont
  {S{\"o}llner}, \citenamefont {Javadi}, \citenamefont {Hansen}, \citenamefont
  {Mahmoodian}, \citenamefont {Liu}, \citenamefont {Thyrrestrup}, \citenamefont
  {Lee}, \citenamefont {Song}, \citenamefont {Stobbe} \emph
  {et~al.}}]{arcari2014near}%
  \BibitemOpen
  \bibfield  {author} {\bibinfo {author} {\bibnamefont {Arcari}, \bibfnamefont
  {M.}}, \bibinfo {author} {\bibnamefont {S{\"o}llner}, \bibfnamefont {I.}},
  \bibinfo {author} {\bibnamefont {Javadi}, \bibfnamefont {A.}}, \bibinfo
  {author} {\bibnamefont {Hansen}, \bibfnamefont {S.~L.}}, \bibinfo {author}
  {\bibnamefont {Mahmoodian}, \bibfnamefont {S.}}, \bibinfo {author}
  {\bibnamefont {Liu}, \bibfnamefont {J.}}, \bibinfo {author} {\bibnamefont
  {Thyrrestrup}, \bibfnamefont {H.}}, \bibinfo {author} {\bibnamefont {Lee},
  \bibfnamefont {E.~H.}}, \bibinfo {author} {\bibnamefont {Song}, \bibfnamefont
  {J.~D.}}, \bibinfo {author} {\bibnamefont {Stobbe}, \bibfnamefont {S.}},
  \emph {et~al.},\ }\href@noop {} {\bibfield  {journal} {\bibinfo  {journal}
  {Physical review letters}\ }\textbf {\bibinfo {volume} {113}},\ \bibinfo
  {pages} {093603} (\bibinfo {year} {2014}{\natexlab{a}})}\BibitemShut
  {NoStop}%
\bibitem [{\citenamefont {Arcari}\ \emph
  {et~al.}(2014{\natexlab{b}})\citenamefont {Arcari}, \citenamefont
  {S\"ollner}, \citenamefont {Javadi}, \citenamefont {Lindskov~Hansen},
  \citenamefont {Mahmoodian}, \citenamefont {Liu}, \citenamefont {Thyrrestrup},
  \citenamefont {Lee}, \citenamefont {Song}, \citenamefont {Stobbe},\ and\
  \citenamefont {Lodahl}}]{PhysRevLett.113.093603}%
  \BibitemOpen
  \bibfield  {author} {\bibinfo {author} {\bibnamefont {Arcari}, \bibfnamefont
  {M.}}, \bibinfo {author} {\bibnamefont {S\"ollner}, \bibfnamefont {I.}},
  \bibinfo {author} {\bibnamefont {Javadi}, \bibfnamefont {A.}}, \bibinfo
  {author} {\bibnamefont {Lindskov~Hansen}, \bibfnamefont {S.}}, \bibinfo
  {author} {\bibnamefont {Mahmoodian}, \bibfnamefont {S.}}, \bibinfo {author}
  {\bibnamefont {Liu}, \bibfnamefont {J.}}, \bibinfo {author} {\bibnamefont
  {Thyrrestrup}, \bibfnamefont {H.}}, \bibinfo {author} {\bibnamefont {Lee},
  \bibfnamefont {E.~H.}}, \bibinfo {author} {\bibnamefont {Song}, \bibfnamefont
  {J.~D.}}, \bibinfo {author} {\bibnamefont {Stobbe}, \bibfnamefont {S.}}, \
  and\ \bibinfo {author} {\bibnamefont {Lodahl}, \bibfnamefont {P.}},\ }\href
  {\doibase 10.1103/PhysRevLett.113.093603} {\bibfield  {journal} {\bibinfo
  {journal} {Phys. Rev. Lett.}\ }\textbf {\bibinfo {volume} {113}},\ \bibinfo
  {pages} {093603} (\bibinfo {year} {2014}{\natexlab{b}})}\BibitemShut
  {NoStop}%
\bibitem [{\citenamefont {Arora}\ \emph {et~al.}(2021)\citenamefont {Arora},
  \citenamefont {Bauer}, \citenamefont {Barczyk}, \citenamefont {Verhagen},\
  and\ \citenamefont {Kuipers}}]{arora2021direct}%
  \BibitemOpen
  \bibfield  {author} {\bibinfo {author} {\bibnamefont {Arora}, \bibfnamefont
  {S.}}, \bibinfo {author} {\bibnamefont {Bauer}, \bibfnamefont {T.}}, \bibinfo
  {author} {\bibnamefont {Barczyk}, \bibfnamefont {R.}}, \bibinfo {author}
  {\bibnamefont {Verhagen}, \bibfnamefont {E.}}, \ and\ \bibinfo {author}
  {\bibnamefont {Kuipers}, \bibfnamefont {L.}},\ }\href@noop {} {\bibfield
  {journal} {\bibinfo  {journal} {Light: Science \& Applications}\ }\textbf
  {\bibinfo {volume} {10}},\ \bibinfo {pages} {1} (\bibinfo {year}
  {2021})}\BibitemShut {NoStop}%
\bibitem [{\citenamefont {Arregui}\ \emph
  {et~al.}(2021{\natexlab{a}})\citenamefont {Arregui}, \citenamefont
  {Gomis-Bresco}, \citenamefont {Sotomayor-Torres},\ and\ \citenamefont
  {Garcia}}]{PhysRevLett.126.027403}%
  \BibitemOpen
  \bibfield  {author} {\bibinfo {author} {\bibnamefont {Arregui}, \bibfnamefont
  {G.}}, \bibinfo {author} {\bibnamefont {Gomis-Bresco}, \bibfnamefont {J.}},
  \bibinfo {author} {\bibnamefont {Sotomayor-Torres}, \bibfnamefont {C.~M.}}, \
  and\ \bibinfo {author} {\bibnamefont {Garcia}, \bibfnamefont {P.~D.}},\
  }\href {\doibase 10.1103/PhysRevLett.126.027403} {\bibfield  {journal}
  {\bibinfo  {journal} {Phys. Rev. Lett.}\ }\textbf {\bibinfo {volume} {126}},\
  \bibinfo {pages} {027403} (\bibinfo {year} {2021}{\natexlab{a}})}\BibitemShut
  {NoStop}%
\bibitem [{\citenamefont {Arregui}\ \emph
  {et~al.}(2021{\natexlab{b}})\citenamefont {Arregui}, \citenamefont
  {Gomis-Bresco}, \citenamefont {Sotomayor-Torres},\ and\ \citenamefont
  {Garcia}}]{arregui2020quantifying}%
  \BibitemOpen
  \bibfield  {author} {\bibinfo {author} {\bibnamefont {Arregui}, \bibfnamefont
  {G.}}, \bibinfo {author} {\bibnamefont {Gomis-Bresco}, \bibfnamefont {J.}},
  \bibinfo {author} {\bibnamefont {Sotomayor-Torres}, \bibfnamefont {C.~M.}}, \
  and\ \bibinfo {author} {\bibnamefont {Garcia}, \bibfnamefont {P.~D.}},\
  }\href@noop {} {\bibfield  {journal} {\bibinfo  {journal} {Physical Review
  Letters}\ }\textbf {\bibinfo {volume} {126}},\ \bibinfo {pages} {027403}
  (\bibinfo {year} {2021}{\natexlab{b}})}\BibitemShut {NoStop}%
\bibitem [{\citenamefont {Barik}\ \emph {et~al.}(2018)\citenamefont {Barik},
  \citenamefont {Karasahin}, \citenamefont {Flower}, \citenamefont {Cai},
  \citenamefont {Miyake}, \citenamefont {DeGottardi}, \citenamefont {Hafezi},\
  and\ \citenamefont {Waks}}]{barik2018topological}%
  \BibitemOpen
  \bibfield  {author} {\bibinfo {author} {\bibnamefont {Barik}, \bibfnamefont
  {S.}}, \bibinfo {author} {\bibnamefont {Karasahin}, \bibfnamefont {A.}},
  \bibinfo {author} {\bibnamefont {Flower}, \bibfnamefont {C.}}, \bibinfo
  {author} {\bibnamefont {Cai}, \bibfnamefont {T.}}, \bibinfo {author}
  {\bibnamefont {Miyake}, \bibfnamefont {H.}}, \bibinfo {author} {\bibnamefont
  {DeGottardi}, \bibfnamefont {W.}}, \bibinfo {author} {\bibnamefont {Hafezi},
  \bibfnamefont {M.}}, \ and\ \bibinfo {author} {\bibnamefont {Waks},
  \bibfnamefont {E.}},\ }\href@noop {} {\bibfield  {journal} {\bibinfo
  {journal} {Science}\ }\textbf {\bibinfo {volume} {359}},\ \bibinfo {pages}
  {666} (\bibinfo {year} {2018})}\BibitemShut {NoStop}%
\bibitem [{\citenamefont {Barik}\ \emph {et~al.}(2020)\citenamefont {Barik},
  \citenamefont {Karasahin}, \citenamefont {Mittal}, \citenamefont {Waks},\
  and\ \citenamefont {Hafezi}}]{barik2020chiral}%
  \BibitemOpen
  \bibfield  {author} {\bibinfo {author} {\bibnamefont {Barik}, \bibfnamefont
  {S.}}, \bibinfo {author} {\bibnamefont {Karasahin}, \bibfnamefont {A.}},
  \bibinfo {author} {\bibnamefont {Mittal}, \bibfnamefont {S.}}, \bibinfo
  {author} {\bibnamefont {Waks}, \bibfnamefont {E.}}, \ and\ \bibinfo {author}
  {\bibnamefont {Hafezi}, \bibfnamefont {M.}},\ }\href@noop {} {\bibfield
  {journal} {\bibinfo  {journal} {Physical Review B}\ }\textbf {\bibinfo
  {volume} {101}},\ \bibinfo {pages} {205303} (\bibinfo {year}
  {2020})}\BibitemShut {NoStop}%
\bibitem [{\citenamefont {Blanco-Redondo}\ \emph {et~al.}(2018)\citenamefont
  {Blanco-Redondo}, \citenamefont {Bell}, \citenamefont {Oren}, \citenamefont
  {Eggleton},\ and\ \citenamefont {Segev}}]{blanco2018topological}%
  \BibitemOpen
  \bibfield  {author} {\bibinfo {author} {\bibnamefont {Blanco-Redondo},
  \bibfnamefont {A.}}, \bibinfo {author} {\bibnamefont {Bell}, \bibfnamefont
  {B.}}, \bibinfo {author} {\bibnamefont {Oren}, \bibfnamefont {D.}}, \bibinfo
  {author} {\bibnamefont {Eggleton}, \bibfnamefont {B.~J.}}, \ and\ \bibinfo
  {author} {\bibnamefont {Segev}, \bibfnamefont {M.}},\ }\href@noop {}
  {\bibfield  {journal} {\bibinfo  {journal} {Science}\ }\textbf {\bibinfo
  {volume} {362}},\ \bibinfo {pages} {568} (\bibinfo {year}
  {2018})}\BibitemShut {NoStop}%
\bibitem [{\citenamefont {Chang}, \citenamefont {Vuleti{\'c}},\ and\
  \citenamefont {Lukin}(2014)}]{chang2014quantum}%
  \BibitemOpen
  \bibfield  {author} {\bibinfo {author} {\bibnamefont {Chang}, \bibfnamefont
  {D.~E.}}, \bibinfo {author} {\bibnamefont {Vuleti{\'c}}, \bibfnamefont {V.}},
  \ and\ \bibinfo {author} {\bibnamefont {Lukin}, \bibfnamefont {M.~D.}},\
  }\href@noop {} {\bibfield  {journal} {\bibinfo  {journal} {Nature Photonics}\
  }\textbf {\bibinfo {volume} {8}},\ \bibinfo {pages} {685} (\bibinfo {year}
  {2014})}\BibitemShut {NoStop}%
\bibitem [{\citenamefont {Christiansen}, \citenamefont {Wang},\ and\
  \citenamefont {Sigmund}(2019)}]{christiansen2019topological}%
  \BibitemOpen
  \bibfield  {author} {\bibinfo {author} {\bibnamefont {Christiansen},
  \bibfnamefont {R.~E.}}, \bibinfo {author} {\bibnamefont {Wang}, \bibfnamefont
  {F.}}, \ and\ \bibinfo {author} {\bibnamefont {Sigmund}, \bibfnamefont
  {O.}},\ }\href@noop {} {\bibfield  {journal} {\bibinfo  {journal} {Physical
  review letters}\ }\textbf {\bibinfo {volume} {122}},\ \bibinfo {pages}
  {234502} (\bibinfo {year} {2019})}\BibitemShut {NoStop}%
\bibitem [{\citenamefont {Christiansen}\ \emph {et~al.}(2019)\citenamefont
  {Christiansen}, \citenamefont {Wang}, \citenamefont {Sigmund},\ and\
  \citenamefont {Stobbe}}]{christiansen2019designing}%
  \BibitemOpen
  \bibfield  {author} {\bibinfo {author} {\bibnamefont {Christiansen},
  \bibfnamefont {R.~E.}}, \bibinfo {author} {\bibnamefont {Wang}, \bibfnamefont
  {F.}}, \bibinfo {author} {\bibnamefont {Sigmund}, \bibfnamefont {O.}}, \ and\
  \bibinfo {author} {\bibnamefont {Stobbe}, \bibfnamefont {S.}},\ }\href@noop
  {} {\bibfield  {journal} {\bibinfo  {journal} {Nanophotonics}\ }\textbf
  {\bibinfo {volume} {8}},\ \bibinfo {pages} {1363} (\bibinfo {year}
  {2019})}\BibitemShut {NoStop}%
\bibitem [{\citenamefont {Coles}\ \emph {et~al.}(2017)\citenamefont {Coles},
  \citenamefont {Price}, \citenamefont {Royall}, \citenamefont {Clarke},
  \citenamefont {Skolnick}, \citenamefont {Fox},\ and\ \citenamefont
  {Makhonin}}]{coles2017path}%
  \BibitemOpen
  \bibfield  {author} {\bibinfo {author} {\bibnamefont {Coles}, \bibfnamefont
  {R.~J.}}, \bibinfo {author} {\bibnamefont {Price}, \bibfnamefont {D.~M.}},
  \bibinfo {author} {\bibnamefont {Royall}, \bibfnamefont {B.}}, \bibinfo
  {author} {\bibnamefont {Clarke}, \bibfnamefont {E.}}, \bibinfo {author}
  {\bibnamefont {Skolnick}, \bibfnamefont {M.~S.}}, \bibinfo {author}
  {\bibnamefont {Fox}, \bibfnamefont {A.~M.}}, \ and\ \bibinfo {author}
  {\bibnamefont {Makhonin}, \bibfnamefont {M.}},\ }\href@noop {} {\bibfield
  {journal} {\bibinfo  {journal} {Physical review B}\ }\textbf {\bibinfo
  {volume} {95}},\ \bibinfo {pages} {121401} (\bibinfo {year}
  {2017})}\BibitemShut {NoStop}%
\bibitem [{\citenamefont {Crane}\ \emph {et~al.}(2017)\citenamefont {Crane},
  \citenamefont {Trojak}, \citenamefont {Vasco}, \citenamefont {Hughes},\ and\
  \citenamefont {Sapienza}}]{crane2017anderson}%
  \BibitemOpen
  \bibfield  {author} {\bibinfo {author} {\bibnamefont {Crane}, \bibfnamefont
  {T.}}, \bibinfo {author} {\bibnamefont {Trojak}, \bibfnamefont {O.~J.}},
  \bibinfo {author} {\bibnamefont {Vasco}, \bibfnamefont {J.~P.}}, \bibinfo
  {author} {\bibnamefont {Hughes}, \bibfnamefont {S.}}, \ and\ \bibinfo
  {author} {\bibnamefont {Sapienza}, \bibfnamefont {L.}},\ }\href@noop {}
  {\bibfield  {journal} {\bibinfo  {journal} {ACS Photonics}\ }\textbf
  {\bibinfo {volume} {4}},\ \bibinfo {pages} {2274} (\bibinfo {year}
  {2017})}\BibitemShut {NoStop}%
\bibitem [{\citenamefont {Ezawa}(2008)}]{ezawa2008quantum}%
  \BibitemOpen
  \bibfield  {author} {\bibinfo {author} {\bibnamefont {Ezawa}, \bibfnamefont
  {Z.~F.}},\ }\href@noop {} {\emph {\bibinfo {title} {Quantum Hall effects:
  Field theoretical approach and related topics}}}\ (\bibinfo  {publisher}
  {World Scientific Publishing Company},\ \bibinfo {year} {2008})\BibitemShut
  {NoStop}%
\bibitem [{\citenamefont {Fang}, \citenamefont {Yu},\ and\ \citenamefont
  {Fan}(2012)}]{fang2012realizing}%
  \BibitemOpen
  \bibfield  {author} {\bibinfo {author} {\bibnamefont {Fang}, \bibfnamefont
  {K.}}, \bibinfo {author} {\bibnamefont {Yu}, \bibfnamefont {Z.}}, \ and\
  \bibinfo {author} {\bibnamefont {Fan}, \bibfnamefont {S.}},\ }\href@noop {}
  {\bibfield  {journal} {\bibinfo  {journal} {Nature photonics}\ }\textbf
  {\bibinfo {volume} {6}},\ \bibinfo {pages} {782} (\bibinfo {year}
  {2012})}\BibitemShut {NoStop}%
\bibitem [{\citenamefont {Gmeiner}\ \emph {et~al.}(2016)\citenamefont
  {Gmeiner}, \citenamefont {Maser}, \citenamefont {Utikal}, \citenamefont
  {G{\"o}tzinger},\ and\ \citenamefont {Sandoghdar}}]{gmeiner2016spectroscopy}%
  \BibitemOpen
  \bibfield  {author} {\bibinfo {author} {\bibnamefont {Gmeiner}, \bibfnamefont
  {B.}}, \bibinfo {author} {\bibnamefont {Maser}, \bibfnamefont {A.}}, \bibinfo
  {author} {\bibnamefont {Utikal}, \bibfnamefont {T.}}, \bibinfo {author}
  {\bibnamefont {G{\"o}tzinger}, \bibfnamefont {S.}}, \ and\ \bibinfo {author}
  {\bibnamefont {Sandoghdar}, \bibfnamefont {V.}},\ }\href@noop {} {\bibfield
  {journal} {\bibinfo  {journal} {Physical Chemistry Chemical Physics}\
  }\textbf {\bibinfo {volume} {18}},\ \bibinfo {pages} {19588} (\bibinfo {year}
  {2016})}\BibitemShut {NoStop}%
\bibitem [{\citenamefont {Gschrey}\ \emph {et~al.}(2015)\citenamefont
  {Gschrey}, \citenamefont {Schmidt}, \citenamefont {Schulze}, \citenamefont
  {Strittmatter}, \citenamefont {Rodt},\ and\ \citenamefont
  {Reitzenstein}}]{gschrey2015resolution}%
  \BibitemOpen
  \bibfield  {author} {\bibinfo {author} {\bibnamefont {Gschrey}, \bibfnamefont
  {M.}}, \bibinfo {author} {\bibnamefont {Schmidt}, \bibfnamefont {R.}},
  \bibinfo {author} {\bibnamefont {Schulze}, \bibfnamefont {J.-H.}}, \bibinfo
  {author} {\bibnamefont {Strittmatter}, \bibfnamefont {A.}}, \bibinfo {author}
  {\bibnamefont {Rodt}, \bibfnamefont {S.}}, \ and\ \bibinfo {author}
  {\bibnamefont {Reitzenstein}, \bibfnamefont {S.}},\ }\href@noop {} {\bibfield
   {journal} {\bibinfo  {journal} {Journal of Vacuum Science \& Technology B,
  Nanotechnology and Microelectronics: Materials, Processing, Measurement, and
  Phenomena}\ }\textbf {\bibinfo {volume} {33}},\ \bibinfo {pages} {021603}
  (\bibinfo {year} {2015})}\BibitemShut {NoStop}%
\bibitem [{\citenamefont {Hafezi}\ \emph {et~al.}(2011)\citenamefont {Hafezi},
  \citenamefont {Demler}, \citenamefont {Lukin},\ and\ \citenamefont
  {Taylor}}]{Hafezi_2011}%
  \BibitemOpen
  \bibfield  {author} {\bibinfo {author} {\bibnamefont {Hafezi}, \bibfnamefont
  {M.}}, \bibinfo {author} {\bibnamefont {Demler}, \bibfnamefont {E.~A.}},
  \bibinfo {author} {\bibnamefont {Lukin}, \bibfnamefont {M.~D.}}, \ and\
  \bibinfo {author} {\bibnamefont {Taylor}, \bibfnamefont {J.~M.}},\ }\href
  {\doibase 10.1038/nphys2063} {\bibfield  {journal} {\bibinfo  {journal}
  {Nature Physics}\ }\textbf {\bibinfo {volume} {7}},\ \bibinfo {pages}
  {907–912} (\bibinfo {year} {2011})}\BibitemShut {NoStop}%
\bibitem [{\citenamefont {Haldane}\ and\ \citenamefont
  {Raghu}(2008)}]{PhysRevLett.100.013904}%
  \BibitemOpen
  \bibfield  {author} {\bibinfo {author} {\bibnamefont {Haldane}, \bibfnamefont
  {F.~D.~M.}}\ and\ \bibinfo {author} {\bibnamefont {Raghu}, \bibfnamefont
  {S.}},\ }\href {\doibase 10.1103/PhysRevLett.100.013904} {\bibfield
  {journal} {\bibinfo  {journal} {Phys. Rev. Lett.}\ }\textbf {\bibinfo
  {volume} {100}},\ \bibinfo {pages} {013904} (\bibinfo {year}
  {2008})}\BibitemShut {NoStop}%
\bibitem [{\citenamefont {Hasan}\ and\ \citenamefont
  {Kane}(2010)}]{RevModPhys.82.3045}%
  \BibitemOpen
  \bibfield  {author} {\bibinfo {author} {\bibnamefont {Hasan}, \bibfnamefont
  {M.~Z.}}\ and\ \bibinfo {author} {\bibnamefont {Kane}, \bibfnamefont
  {C.~L.}},\ }\href {\doibase 10.1103/RevModPhys.82.3045} {\bibfield  {journal}
  {\bibinfo  {journal} {Rev. Mod. Phys.}\ }\textbf {\bibinfo {volume} {82}},\
  \bibinfo {pages} {3045} (\bibinfo {year} {2010})}\BibitemShut {NoStop}%
\bibitem [{\citenamefont {He}\ \emph {et~al.}(2019)\citenamefont {He},
  \citenamefont {Liang}, \citenamefont {Yuan}, \citenamefont {Qiu},
  \citenamefont {Chen}, \citenamefont {Zhao},\ and\ \citenamefont
  {Dong}}]{he2019silicon}%
  \BibitemOpen
  \bibfield  {author} {\bibinfo {author} {\bibnamefont {He}, \bibfnamefont
  {X.-T.}}, \bibinfo {author} {\bibnamefont {Liang}, \bibfnamefont {E.-T.}},
  \bibinfo {author} {\bibnamefont {Yuan}, \bibfnamefont {J.-J.}}, \bibinfo
  {author} {\bibnamefont {Qiu}, \bibfnamefont {H.-Y.}}, \bibinfo {author}
  {\bibnamefont {Chen}, \bibfnamefont {X.-D.}}, \bibinfo {author} {\bibnamefont
  {Zhao}, \bibfnamefont {F.-L.}}, \ and\ \bibinfo {author} {\bibnamefont
  {Dong}, \bibfnamefont {J.-W.}},\ }\href@noop {} {\bibfield  {journal}
  {\bibinfo  {journal} {Nature communications}\ }\textbf {\bibinfo {volume}
  {10}},\ \bibinfo {pages} {1} (\bibinfo {year} {2019})}\BibitemShut {NoStop}%
\bibitem [{\citenamefont {Hughes}(2004)}]{hughes2004enhanced}%
  \BibitemOpen
  \bibfield  {author} {\bibinfo {author} {\bibnamefont {Hughes}, \bibfnamefont
  {S.}},\ }\href@noop {} {\bibfield  {journal} {\bibinfo  {journal} {Optics
  letters}\ }\textbf {\bibinfo {volume} {29}},\ \bibinfo {pages} {2659}
  (\bibinfo {year} {2004})}\BibitemShut {NoStop}%
\bibitem [{\citenamefont {Hughes}\ \emph {et~al.}(2005)\citenamefont {Hughes},
  \citenamefont {Ramunno}, \citenamefont {Young},\ and\ \citenamefont
  {Sipe}}]{PhysRevLett.94.033903}%
  \BibitemOpen
  \bibfield  {author} {\bibinfo {author} {\bibnamefont {Hughes}, \bibfnamefont
  {S.}}, \bibinfo {author} {\bibnamefont {Ramunno}, \bibfnamefont {L.}},
  \bibinfo {author} {\bibnamefont {Young}, \bibfnamefont {J.~F.}}, \ and\
  \bibinfo {author} {\bibnamefont {Sipe}, \bibfnamefont {J.~E.}},\ }\href
  {\doibase 10.1103/PhysRevLett.94.033903} {\bibfield  {journal} {\bibinfo
  {journal} {Phys. Rev. Lett.}\ }\textbf {\bibinfo {volume} {94}},\ \bibinfo
  {pages} {033903} (\bibinfo {year} {2005})}\BibitemShut {NoStop}%
\bibitem [{\citenamefont {Husko}\ \emph {et~al.}(2016)\citenamefont {Husko},
  \citenamefont {Wulf}, \citenamefont {Lefrancois}, \citenamefont
  {Combri{\'e}}, \citenamefont {Lehoucq}, \citenamefont {De~Rossi},
  \citenamefont {Eggleton},\ and\ \citenamefont {Kuipers}}]{husko2016free}%
  \BibitemOpen
  \bibfield  {author} {\bibinfo {author} {\bibnamefont {Husko}, \bibfnamefont
  {C.}}, \bibinfo {author} {\bibnamefont {Wulf}, \bibfnamefont {M.}}, \bibinfo
  {author} {\bibnamefont {Lefrancois}, \bibfnamefont {S.}}, \bibinfo {author}
  {\bibnamefont {Combri{\'e}}, \bibfnamefont {S.}}, \bibinfo {author}
  {\bibnamefont {Lehoucq}, \bibfnamefont {G.}}, \bibinfo {author} {\bibnamefont
  {De~Rossi}, \bibfnamefont {A.}}, \bibinfo {author} {\bibnamefont {Eggleton},
  \bibfnamefont {B.~J.}}, \ and\ \bibinfo {author} {\bibnamefont {Kuipers},
  \bibfnamefont {L.}},\ }\href@noop {} {\bibfield  {journal} {\bibinfo
  {journal} {Nature Communications}\ }\textbf {\bibinfo {volume} {7}},\
  \bibinfo {pages} {1} (\bibinfo {year} {2016})}\BibitemShut {NoStop}%
\bibitem [{\citenamefont {Jalali~Mehrabad}\ \emph {et~al.}(2020)\citenamefont
  {Jalali~Mehrabad}, \citenamefont {Foster}, \citenamefont {Dost},
  \citenamefont {Clarke}, \citenamefont {Patil}, \citenamefont {Farrer},
  \citenamefont {Heffernan}, \citenamefont {Skolnick},\ and\ \citenamefont
  {Wilson}}]{jalali2020semiconductor}%
  \BibitemOpen
  \bibfield  {author} {\bibinfo {author} {\bibnamefont {Jalali~Mehrabad},
  \bibfnamefont {M.}}, \bibinfo {author} {\bibnamefont {Foster}, \bibfnamefont
  {A.}}, \bibinfo {author} {\bibnamefont {Dost}, \bibfnamefont {R.}}, \bibinfo
  {author} {\bibnamefont {Clarke}, \bibfnamefont {E.}}, \bibinfo {author}
  {\bibnamefont {Patil}, \bibfnamefont {P.}}, \bibinfo {author} {\bibnamefont
  {Farrer}, \bibfnamefont {I.}}, \bibinfo {author} {\bibnamefont {Heffernan},
  \bibfnamefont {J.}}, \bibinfo {author} {\bibnamefont {Skolnick},
  \bibfnamefont {M.}}, \ and\ \bibinfo {author} {\bibnamefont {Wilson},
  \bibfnamefont {L.}},\ }\href@noop {} {\bibfield  {journal} {\bibinfo
  {journal} {Applied Physics Letters}\ }\textbf {\bibinfo {volume} {116}},\
  \bibinfo {pages} {061102} (\bibinfo {year} {2020})}\BibitemShut {NoStop}%
\bibitem [{\citenamefont {Javadi}\ \emph {et~al.}(2015)\citenamefont {Javadi},
  \citenamefont {S{\"o}llner}, \citenamefont {Arcari}, \citenamefont {Hansen},
  \citenamefont {Midolo}, \citenamefont {Mahmoodian}, \citenamefont
  {Kir{\v{s}}ansk{\.e}}, \citenamefont {Pregnolato}, \citenamefont {Lee},
  \citenamefont {Song} \emph {et~al.}}]{javadi2015single}%
  \BibitemOpen
  \bibfield  {author} {\bibinfo {author} {\bibnamefont {Javadi}, \bibfnamefont
  {A.}}, \bibinfo {author} {\bibnamefont {S{\"o}llner}, \bibfnamefont {I.}},
  \bibinfo {author} {\bibnamefont {Arcari}, \bibfnamefont {M.}}, \bibinfo
  {author} {\bibnamefont {Hansen}, \bibfnamefont {S.~L.}}, \bibinfo {author}
  {\bibnamefont {Midolo}, \bibfnamefont {L.}}, \bibinfo {author} {\bibnamefont
  {Mahmoodian}, \bibfnamefont {S.}}, \bibinfo {author} {\bibnamefont
  {Kir{\v{s}}ansk{\.e}}, \bibfnamefont {G.}}, \bibinfo {author} {\bibnamefont
  {Pregnolato}, \bibfnamefont {T.}}, \bibinfo {author} {\bibnamefont {Lee},
  \bibfnamefont {E.}}, \bibinfo {author} {\bibnamefont {Song}, \bibfnamefont
  {J.}},  \emph {et~al.},\ }\href@noop {} {\bibfield  {journal} {\bibinfo
  {journal} {Nature communications}\ }\textbf {\bibinfo {volume} {6}},\
  \bibinfo {pages} {1} (\bibinfo {year} {2015})}\BibitemShut {NoStop}%
\bibitem [{\citenamefont {Jin}\ \emph {et~al.}(2019)\citenamefont {Jin},
  \citenamefont {Yin}, \citenamefont {Ni}, \citenamefont {Solja{\v{c}}i{\'c}},
  \citenamefont {Zhen},\ and\ \citenamefont {Peng}}]{jin2019topologically}%
  \BibitemOpen
  \bibfield  {author} {\bibinfo {author} {\bibnamefont {Jin}, \bibfnamefont
  {J.}}, \bibinfo {author} {\bibnamefont {Yin}, \bibfnamefont {X.}}, \bibinfo
  {author} {\bibnamefont {Ni}, \bibfnamefont {L.}}, \bibinfo {author}
  {\bibnamefont {Solja{\v{c}}i{\'c}}, \bibfnamefont {M.}}, \bibinfo {author}
  {\bibnamefont {Zhen}, \bibfnamefont {B.}}, \ and\ \bibinfo {author}
  {\bibnamefont {Peng}, \bibfnamefont {C.}},\ }\href@noop {} {\bibfield
  {journal} {\bibinfo  {journal} {Nature}\ }\textbf {\bibinfo {volume} {574}},\
  \bibinfo {pages} {501} (\bibinfo {year} {2019})}\BibitemShut {NoStop}%
\bibitem [{\citenamefont {Joannopoulos}\ \emph {et~al.}(2008)\citenamefont
  {Joannopoulos}, \citenamefont {Johnson}, \citenamefont {Winn},\ and\
  \citenamefont {Meade}}]{joannopoulos2008molding}%
  \BibitemOpen
  \bibfield  {author} {\bibinfo {author} {\bibnamefont {Joannopoulos},
  \bibfnamefont {J.~D.}}, \bibinfo {author} {\bibnamefont {Johnson},
  \bibfnamefont {S.~G.}}, \bibinfo {author} {\bibnamefont {Winn}, \bibfnamefont
  {J.~N.}}, \ and\ \bibinfo {author} {\bibnamefont {Meade}, \bibfnamefont
  {R.~D.}},\ }\href@noop {} {\bibfield  {journal} {\bibinfo  {journal}
  {Princeton Univ. Press, Princeton, NJ [ua]}\ } (\bibinfo {year}
  {2008})}\BibitemShut {NoStop}%
\bibitem [{\citenamefont {Joannopoulos}, \citenamefont {Villeneuve},\ and\
  \citenamefont {Fan}(1997)}]{joannopoulos1997photonic}%
  \BibitemOpen
  \bibfield  {author} {\bibinfo {author} {\bibnamefont {Joannopoulos},
  \bibfnamefont {J.~D.}}, \bibinfo {author} {\bibnamefont {Villeneuve},
  \bibfnamefont {P.~R.}}, \ and\ \bibinfo {author} {\bibnamefont {Fan},
  \bibfnamefont {S.}},\ }\href@noop {} {\bibfield  {journal} {\bibinfo
  {journal} {Solid State Communications}\ }\textbf {\bibinfo {volume} {102}},\
  \bibinfo {pages} {165} (\bibinfo {year} {1997})}\BibitemShut {NoStop}%
\bibitem [{\citenamefont {Khanikaev}\ \emph {et~al.}(2013)\citenamefont
  {Khanikaev}, \citenamefont {Mousavi}, \citenamefont {Tse}, \citenamefont
  {Kargarian}, \citenamefont {MacDonald},\ and\ \citenamefont
  {Shvets}}]{khanikaev2013photonic}%
  \BibitemOpen
  \bibfield  {author} {\bibinfo {author} {\bibnamefont {Khanikaev},
  \bibfnamefont {A.~B.}}, \bibinfo {author} {\bibnamefont {Mousavi},
  \bibfnamefont {S.~H.}}, \bibinfo {author} {\bibnamefont {Tse}, \bibfnamefont
  {W.-K.}}, \bibinfo {author} {\bibnamefont {Kargarian}, \bibfnamefont {M.}},
  \bibinfo {author} {\bibnamefont {MacDonald}, \bibfnamefont {A.~H.}}, \ and\
  \bibinfo {author} {\bibnamefont {Shvets}, \bibfnamefont {G.}},\ }\href@noop
  {} {\bibfield  {journal} {\bibinfo  {journal} {Nature materials}\ }\textbf
  {\bibinfo {volume} {12}},\ \bibinfo {pages} {233} (\bibinfo {year}
  {2013})}\BibitemShut {NoStop}%
\bibitem [{\citenamefont {Koshino}, \citenamefont {Ishizaka},\ and\
  \citenamefont {Nakamura}(2010)}]{koshino2010deterministic}%
  \BibitemOpen
  \bibfield  {author} {\bibinfo {author} {\bibnamefont {Koshino}, \bibfnamefont
  {K.}}, \bibinfo {author} {\bibnamefont {Ishizaka}, \bibfnamefont {S.}}, \
  and\ \bibinfo {author} {\bibnamefont {Nakamura}, \bibfnamefont {Y.}},\
  }\href@noop {} {\bibfield  {journal} {\bibinfo  {journal} {Physical Review
  A}\ }\textbf {\bibinfo {volume} {82}},\ \bibinfo {pages} {010301} (\bibinfo
  {year} {2010})}\BibitemShut {NoStop}%
\bibitem [{\citenamefont {Kuramochi}\ \emph {et~al.}(2005)\citenamefont
  {Kuramochi}, \citenamefont {Notomi}, \citenamefont {Hughes}, \citenamefont
  {Shinya}, \citenamefont {Watanabe},\ and\ \citenamefont
  {Ramunno}}]{PhysRevB.72.161318}%
  \BibitemOpen
  \bibfield  {author} {\bibinfo {author} {\bibnamefont {Kuramochi},
  \bibfnamefont {E.}}, \bibinfo {author} {\bibnamefont {Notomi}, \bibfnamefont
  {M.}}, \bibinfo {author} {\bibnamefont {Hughes}, \bibfnamefont {S.}},
  \bibinfo {author} {\bibnamefont {Shinya}, \bibfnamefont {A.}}, \bibinfo
  {author} {\bibnamefont {Watanabe}, \bibfnamefont {T.}}, \ and\ \bibinfo
  {author} {\bibnamefont {Ramunno}, \bibfnamefont {L.}},\ }\href {\doibase
  10.1103/PhysRevB.72.161318} {\bibfield  {journal} {\bibinfo  {journal} {Phys.
  Rev. B}\ }\textbf {\bibinfo {volume} {72}},\ \bibinfo {pages} {161318}
  (\bibinfo {year} {2005})}\BibitemShut {NoStop}%
\bibitem [{\citenamefont {Le~Feber}, \citenamefont {Rotenberg},\ and\
  \citenamefont {Kuipers}(2015)}]{le2015nanophotonic}%
  \BibitemOpen
  \bibfield  {author} {\bibinfo {author} {\bibnamefont {Le~Feber},
  \bibfnamefont {B.}}, \bibinfo {author} {\bibnamefont {Rotenberg},
  \bibfnamefont {N.}}, \ and\ \bibinfo {author} {\bibnamefont {Kuipers},
  \bibfnamefont {L.}},\ }\href@noop {} {\bibfield  {journal} {\bibinfo
  {journal} {Nature communications}\ }\textbf {\bibinfo {volume} {6}},\
  \bibinfo {pages} {1} (\bibinfo {year} {2015})}\BibitemShut {NoStop}%
\bibitem [{\citenamefont {Li}\ \emph {et~al.}(2009)\citenamefont {Li},
  \citenamefont {Chu}, \citenamefont {Jain},\ and\ \citenamefont
  {Shen}}]{li2009topological}%
  \BibitemOpen
  \bibfield  {author} {\bibinfo {author} {\bibnamefont {Li}, \bibfnamefont
  {J.}}, \bibinfo {author} {\bibnamefont {Chu}, \bibfnamefont {R.-L.}},
  \bibinfo {author} {\bibnamefont {Jain}, \bibfnamefont {J.~K.}}, \ and\
  \bibinfo {author} {\bibnamefont {Shen}, \bibfnamefont {S.-Q.}},\ }\href@noop
  {} {\bibfield  {journal} {\bibinfo  {journal} {Physical review letters}\
  }\textbf {\bibinfo {volume} {102}},\ \bibinfo {pages} {136806} (\bibinfo
  {year} {2009})}\BibitemShut {NoStop}%
\bibitem [{\citenamefont {Lin}\ \emph {et~al.}(2013)\citenamefont {Lin},
  \citenamefont {M\"uller}, \citenamefont {Wang}, \citenamefont {Yuan},
  \citenamefont {Antoniou}, \citenamefont {Yuan},\ and\ \citenamefont
  {Capasso}}]{lin2013polarization}%
  \BibitemOpen
  \bibfield  {author} {\bibinfo {author} {\bibnamefont {Lin}, \bibfnamefont
  {J.}}, \bibinfo {author} {\bibnamefont {M\"uller}, \bibfnamefont {J.~B.}},
  \bibinfo {author} {\bibnamefont {Wang}, \bibfnamefont {Q.}}, \bibinfo
  {author} {\bibnamefont {Yuan}, \bibfnamefont {G.}}, \bibinfo {author}
  {\bibnamefont {Antoniou}, \bibfnamefont {N.}}, \bibinfo {author}
  {\bibnamefont {Yuan}, \bibfnamefont {X.-C.}}, \ and\ \bibinfo {author}
  {\bibnamefont {Capasso}, \bibfnamefont {F.}},\ }\href@noop {} {\bibfield
  {journal} {\bibinfo  {journal} {Science}\ }\textbf {\bibinfo {volume}
  {340}},\ \bibinfo {pages} {331} (\bibinfo {year} {2013})}\BibitemShut
  {NoStop}%
\bibitem [{\citenamefont {Liu}, \citenamefont {Srinivasan},\ and\ \citenamefont
  {Liu}()}]{liunanoscale}%
  \BibitemOpen
  \bibfield  {author} {\bibinfo {author} {\bibnamefont {Liu}, \bibfnamefont
  {S.}}, \bibinfo {author} {\bibnamefont {Srinivasan}, \bibfnamefont {K.}}, \
  and\ \bibinfo {author} {\bibnamefont {Liu}, \bibfnamefont {J.}},\ }\href@noop
  {} {\bibinfo  {journal} {Laser \& Photonics Reviews}\ ,\ \bibinfo {pages}
  {2100223}}\BibitemShut {NoStop}%
\bibitem [{\citenamefont {Lodahl}, \citenamefont {Mahmoodian},\ and\
  \citenamefont {Stobbe}(2015)}]{lodahl2015interfacing}%
  \BibitemOpen
\bibfield  {journal} {  }\bibfield  {author} {\bibinfo {author} {\bibnamefont
  {Lodahl}, \bibfnamefont {P.}}, \bibinfo {author} {\bibnamefont {Mahmoodian},
  \bibfnamefont {S.}}, \ and\ \bibinfo {author} {\bibnamefont {Stobbe},
  \bibfnamefont {S.}},\ }\href@noop {} {\bibfield  {journal} {\bibinfo
  {journal} {Reviews of Modern Physics}\ }\textbf {\bibinfo {volume} {87}},\
  \bibinfo {pages} {347} (\bibinfo {year} {2015})}\BibitemShut {NoStop}%
\bibitem [{\citenamefont {Lodahl}\ \emph {et~al.}(2017)\citenamefont {Lodahl},
  \citenamefont {Mahmoodian}, \citenamefont {Stobbe}, \citenamefont
  {Rauschenbeutel}, \citenamefont {Schneeweiss}, \citenamefont {Volz},
  \citenamefont {Pichler},\ and\ \citenamefont {Zoller}}]{lodahl2017chiral}%
  \BibitemOpen
  \bibfield  {author} {\bibinfo {author} {\bibnamefont {Lodahl}, \bibfnamefont
  {P.}}, \bibinfo {author} {\bibnamefont {Mahmoodian}, \bibfnamefont {S.}},
  \bibinfo {author} {\bibnamefont {Stobbe}, \bibfnamefont {S.}}, \bibinfo
  {author} {\bibnamefont {Rauschenbeutel}, \bibfnamefont {A.}}, \bibinfo
  {author} {\bibnamefont {Schneeweiss}, \bibfnamefont {P.}}, \bibinfo {author}
  {\bibnamefont {Volz}, \bibfnamefont {J.}}, \bibinfo {author} {\bibnamefont
  {Pichler}, \bibfnamefont {H.}}, \ and\ \bibinfo {author} {\bibnamefont
  {Zoller}, \bibfnamefont {P.}},\ }\href@noop {} {\bibfield  {journal}
  {\bibinfo  {journal} {Nature}\ }\textbf {\bibinfo {volume} {541}},\ \bibinfo
  {pages} {473} (\bibinfo {year} {2017})}\BibitemShut {NoStop}%
\bibitem [{\citenamefont {Ma}\ and\ \citenamefont {Shvets}(2016)}]{ma2016all}%
  \BibitemOpen
  \bibfield  {author} {\bibinfo {author} {\bibnamefont {Ma}, \bibfnamefont
  {T.}}\ and\ \bibinfo {author} {\bibnamefont {Shvets}, \bibfnamefont {G.}},\
  }\href@noop {} {\bibfield  {journal} {\bibinfo  {journal} {New Journal of
  Physics}\ }\textbf {\bibinfo {volume} {18}},\ \bibinfo {pages} {025012}
  (\bibinfo {year} {2016})}\BibitemShut {NoStop}%
\bibitem [{\citenamefont {Mahmoodian}, \citenamefont {Lodahl},\ and\
  \citenamefont {S{\o}rensen}(2016)}]{mahmoodian2016quantum}%
  \BibitemOpen
  \bibfield  {author} {\bibinfo {author} {\bibnamefont {Mahmoodian},
  \bibfnamefont {S.}}, \bibinfo {author} {\bibnamefont {Lodahl}, \bibfnamefont
  {P.}}, \ and\ \bibinfo {author} {\bibnamefont {S{\o}rensen}, \bibfnamefont
  {A.~S.}},\ }\href@noop {} {\bibfield  {journal} {\bibinfo  {journal}
  {Physical review letters}\ }\textbf {\bibinfo {volume} {117}},\ \bibinfo
  {pages} {240501} (\bibinfo {year} {2016})}\BibitemShut {NoStop}%
\bibitem [{\citenamefont {Mahmoodian}\ \emph {et~al.}(2017)\citenamefont
  {Mahmoodian}, \citenamefont {Prindal-Nielsen}, \citenamefont {S{\"o}llner},
  \citenamefont {Stobbe},\ and\ \citenamefont
  {Lodahl}}]{mahmoodian2017engineering}%
  \BibitemOpen
  \bibfield  {author} {\bibinfo {author} {\bibnamefont {Mahmoodian},
  \bibfnamefont {S.}}, \bibinfo {author} {\bibnamefont {Prindal-Nielsen},
  \bibfnamefont {K.}}, \bibinfo {author} {\bibnamefont {S{\"o}llner},
  \bibfnamefont {I.}}, \bibinfo {author} {\bibnamefont {Stobbe}, \bibfnamefont
  {S.}}, \ and\ \bibinfo {author} {\bibnamefont {Lodahl}, \bibfnamefont {P.}},\
  }\href@noop {} {\bibfield  {journal} {\bibinfo  {journal} {Optical Materials
  Express}\ }\textbf {\bibinfo {volume} {7}},\ \bibinfo {pages} {43} (\bibinfo
  {year} {2017})}\BibitemShut {NoStop}%
\bibitem [{\citenamefont {Mann}\ and\ \citenamefont
  {Hughes}(2017)}]{PhysRevLett.118.253901}%
  \BibitemOpen
  \bibfield  {author} {\bibinfo {author} {\bibnamefont {Mann}, \bibfnamefont
  {N.}}\ and\ \bibinfo {author} {\bibnamefont {Hughes}, \bibfnamefont {S.}},\
  }\href {\doibase 10.1103/PhysRevLett.118.253901} {\bibfield  {journal}
  {\bibinfo  {journal} {Phys. Rev. Lett.}\ }\textbf {\bibinfo {volume} {118}},\
  \bibinfo {pages} {253901} (\bibinfo {year} {2017})}\BibitemShut {NoStop}%
\bibitem [{\citenamefont {Mann}\ \emph {et~al.}(2015)\citenamefont {Mann},
  \citenamefont {Javadi}, \citenamefont {Garc{\'\i}a}, \citenamefont {Lodahl},\
  and\ \citenamefont {Hughes}}]{mann2015theory}%
  \BibitemOpen
  \bibfield  {author} {\bibinfo {author} {\bibnamefont {Mann}, \bibfnamefont
  {N.}}, \bibinfo {author} {\bibnamefont {Javadi}, \bibfnamefont {A.}},
  \bibinfo {author} {\bibnamefont {Garc{\'\i}a}, \bibfnamefont {P.}}, \bibinfo
  {author} {\bibnamefont {Lodahl}, \bibfnamefont {P.}}, \ and\ \bibinfo
  {author} {\bibnamefont {Hughes}, \bibfnamefont {S.}},\ }\href@noop {}
  {\bibfield  {journal} {\bibinfo  {journal} {Physical Review A}\ }\textbf
  {\bibinfo {volume} {92}},\ \bibinfo {pages} {023849} (\bibinfo {year}
  {2015})}\BibitemShut {NoStop}%
\bibitem [{\citenamefont {Martin-Cano}, \citenamefont {Haakh},\ and\
  \citenamefont {Rotenberg}(2019)}]{martin2019chiral}%
  \BibitemOpen
  \bibfield  {author} {\bibinfo {author} {\bibnamefont {Martin-Cano},
  \bibfnamefont {D.}}, \bibinfo {author} {\bibnamefont {Haakh}, \bibfnamefont
  {H.~R.}}, \ and\ \bibinfo {author} {\bibnamefont {Rotenberg}, \bibfnamefont
  {N.}},\ }\href@noop {} {\bibfield  {journal} {\bibinfo  {journal} {Acs
  Photonics}\ }\textbf {\bibinfo {volume} {6}},\ \bibinfo {pages} {961}
  (\bibinfo {year} {2019})}\BibitemShut {NoStop}%
\bibitem [{\citenamefont {Mehrabad}\ \emph {et~al.}(2020)\citenamefont
  {Mehrabad}, \citenamefont {Foster}, \citenamefont {Dost}, \citenamefont
  {Clarke}, \citenamefont {Patil}, \citenamefont {Fox}, \citenamefont
  {Skolnick},\ and\ \citenamefont {Wilson}}]{mehrabad2020chiral}%
  \BibitemOpen
  \bibfield  {author} {\bibinfo {author} {\bibnamefont {Mehrabad},
  \bibfnamefont {M.~J.}}, \bibinfo {author} {\bibnamefont {Foster},
  \bibfnamefont {A.~P.}}, \bibinfo {author} {\bibnamefont {Dost}, \bibfnamefont
  {R.}}, \bibinfo {author} {\bibnamefont {Clarke}, \bibfnamefont {E.}},
  \bibinfo {author} {\bibnamefont {Patil}, \bibfnamefont {P.~K.}}, \bibinfo
  {author} {\bibnamefont {Fox}, \bibfnamefont {A.~M.}}, \bibinfo {author}
  {\bibnamefont {Skolnick}, \bibfnamefont {M.~S.}}, \ and\ \bibinfo {author}
  {\bibnamefont {Wilson}, \bibfnamefont {L.~R.}},\ }\href@noop {} {\bibfield
  {journal} {\bibinfo  {journal} {Optica}\ }\textbf {\bibinfo {volume} {7}},\
  \bibinfo {pages} {1690} (\bibinfo {year} {2020})}\BibitemShut {NoStop}%
\bibitem [{\citenamefont {Nussbaum}, \citenamefont {Sauer},\ and\ \citenamefont
  {Hughes}(2021)}]{nussbaum2021inverse}%
  \BibitemOpen
  \bibfield  {author} {\bibinfo {author} {\bibnamefont {Nussbaum},
  \bibfnamefont {E.}}, \bibinfo {author} {\bibnamefont {Sauer}, \bibfnamefont
  {E.}}, \ and\ \bibinfo {author} {\bibnamefont {Hughes}, \bibfnamefont {S.}},\
  }\href@noop {} {\bibfield  {journal} {\bibinfo  {journal} {Optics Letters}\
  }\textbf {\bibinfo {volume} {46}},\ \bibinfo {pages} {1732} (\bibinfo {year}
  {2021})}\BibitemShut {NoStop}%
\bibitem [{\citenamefont {Parini}\ \emph {et~al.}(2008)\citenamefont {Parini},
  \citenamefont {Hamel}, \citenamefont {De~Rossi}, \citenamefont {Combri{\'e}},
  \citenamefont {Gottesman}, \citenamefont {Gabet}, \citenamefont {Talneau},
  \citenamefont {Jaou{\"e}n}, \citenamefont {Vadal{\`a}} \emph
  {et~al.}}]{parini2008time}%
  \BibitemOpen
  \bibfield  {author} {\bibinfo {author} {\bibnamefont {Parini}, \bibfnamefont
  {A.}}, \bibinfo {author} {\bibnamefont {Hamel}, \bibfnamefont {P.}}, \bibinfo
  {author} {\bibnamefont {De~Rossi}, \bibfnamefont {A.}}, \bibinfo {author}
  {\bibnamefont {Combri{\'e}}, \bibfnamefont {S.}}, \bibinfo {author}
  {\bibnamefont {Gottesman}, \bibfnamefont {Y.}}, \bibinfo {author}
  {\bibnamefont {Gabet}, \bibfnamefont {R.}}, \bibinfo {author} {\bibnamefont
  {Talneau}, \bibfnamefont {A.}}, \bibinfo {author} {\bibnamefont {Jaou{\"e}n},
  \bibfnamefont {Y.}}, \bibinfo {author} {\bibnamefont {Vadal{\`a}},
  \bibfnamefont {G.}},  \emph {et~al.},\ }\href@noop {} {\bibfield  {journal}
  {\bibinfo  {journal} {Journal of lightwave technology}\ }\textbf {\bibinfo
  {volume} {26}},\ \bibinfo {pages} {3794} (\bibinfo {year}
  {2008})}\BibitemShut {NoStop}%
\bibitem [{\citenamefont {Patterson}\ and\ \citenamefont
  {Hughes}(2010)}]{PhysRevB.81.245321}%
  \BibitemOpen
  \bibfield  {author} {\bibinfo {author} {\bibnamefont {Patterson},
  \bibfnamefont {M.}}\ and\ \bibinfo {author} {\bibnamefont {Hughes},
  \bibfnamefont {S.}},\ }\href {\doibase 10.1103/PhysRevB.81.245321} {\bibfield
   {journal} {\bibinfo  {journal} {Phys. Rev. B}\ }\textbf {\bibinfo {volume}
  {81}},\ \bibinfo {pages} {245321} (\bibinfo {year} {2010})}\BibitemShut
  {NoStop}%
\bibitem [{\citenamefont {Patterson}\ \emph
  {et~al.}(2009{\natexlab{a}})\citenamefont {Patterson}, \citenamefont
  {Hughes}, \citenamefont {Combri{\'e}}, \citenamefont {Tran}, \citenamefont
  {De~Rossi}, \citenamefont {Gabet},\ and\ \citenamefont
  {Jaou{\"e}n}}]{patterson2009disorder}%
  \BibitemOpen
  \bibfield  {author} {\bibinfo {author} {\bibnamefont {Patterson},
  \bibfnamefont {M.}}, \bibinfo {author} {\bibnamefont {Hughes}, \bibfnamefont
  {S.}}, \bibinfo {author} {\bibnamefont {Combri{\'e}}, \bibfnamefont {S.}},
  \bibinfo {author} {\bibnamefont {Tran}, \bibfnamefont {N.-V.-Q.}}, \bibinfo
  {author} {\bibnamefont {De~Rossi}, \bibfnamefont {A.}}, \bibinfo {author}
  {\bibnamefont {Gabet}, \bibfnamefont {R.}}, \ and\ \bibinfo {author}
  {\bibnamefont {Jaou{\"e}n}, \bibfnamefont {Y.}},\ }\href@noop {} {\bibfield
  {journal} {\bibinfo  {journal} {Physical review letters}\ }\textbf {\bibinfo
  {volume} {102}},\ \bibinfo {pages} {253903} (\bibinfo {year}
  {2009}{\natexlab{a}})}\BibitemShut {NoStop}%
\bibitem [{\citenamefont {Patterson}\ \emph
  {et~al.}(2009{\natexlab{b}})\citenamefont {Patterson}, \citenamefont
  {Hughes}, \citenamefont {Schulz}, \citenamefont {Beggs}, \citenamefont
  {White}, \citenamefont {O'Faolain},\ and\ \citenamefont
  {Krauss}}]{PhysRevB.80.195305}%
  \BibitemOpen
  \bibfield  {author} {\bibinfo {author} {\bibnamefont {Patterson},
  \bibfnamefont {M.}}, \bibinfo {author} {\bibnamefont {Hughes}, \bibfnamefont
  {S.}}, \bibinfo {author} {\bibnamefont {Schulz}, \bibfnamefont {S.}},
  \bibinfo {author} {\bibnamefont {Beggs}, \bibfnamefont {D.~M.}}, \bibinfo
  {author} {\bibnamefont {White}, \bibfnamefont {T.~P.}}, \bibinfo {author}
  {\bibnamefont {O'Faolain}, \bibfnamefont {L.}}, \ and\ \bibinfo {author}
  {\bibnamefont {Krauss}, \bibfnamefont {T.~F.}},\ }\href {\doibase
  10.1103/PhysRevB.80.195305} {\bibfield  {journal} {\bibinfo  {journal} {Phys.
  Rev. B}\ }\textbf {\bibinfo {volume} {80}},\ \bibinfo {pages} {195305}
  (\bibinfo {year} {2009}{\natexlab{b}})}\BibitemShut {NoStop}%
\bibitem [{\citenamefont {Pedersen}\ \emph {et~al.}(2021)\citenamefont
  {Pedersen}, \citenamefont {Gonz{\'a}lez-Ruiz}, \citenamefont {Hauff},
  \citenamefont {Wang}, \citenamefont {Wieck}, \citenamefont {Ludwig},
  \citenamefont {Schott}, \citenamefont {Midolo}, \citenamefont {S{\o}rensen},
  \citenamefont {Uppu} \emph {et~al.}}]{pedersen2021demand}%
  \BibitemOpen
  \bibfield  {author} {\bibinfo {author} {\bibnamefont {Pedersen},
  \bibfnamefont {F.~T.}}, \bibinfo {author} {\bibnamefont {Gonz{\'a}lez-Ruiz},
  \bibfnamefont {E.~M.}}, \bibinfo {author} {\bibnamefont {Hauff},
  \bibfnamefont {N.}}, \bibinfo {author} {\bibnamefont {Wang}, \bibfnamefont
  {Y.}}, \bibinfo {author} {\bibnamefont {Wieck}, \bibfnamefont {A.~D.}},
  \bibinfo {author} {\bibnamefont {Ludwig}, \bibfnamefont {A.}}, \bibinfo
  {author} {\bibnamefont {Schott}, \bibfnamefont {R.}}, \bibinfo {author}
  {\bibnamefont {Midolo}, \bibfnamefont {L.}}, \bibinfo {author} {\bibnamefont
  {S{\o}rensen}, \bibfnamefont {A.~S.}}, \bibinfo {author} {\bibnamefont
  {Uppu}, \bibfnamefont {R.}},  \emph {et~al.},\ }\href@noop {} {\bibfield
  {journal} {\bibinfo  {journal} {arXiv preprint arXiv:2109.03519}\ } (\bibinfo
  {year} {2021})}\BibitemShut {NoStop}%
\bibitem [{\citenamefont {Petersen}, \citenamefont {Volz},\ and\ \citenamefont
  {Rauschenbeutel}(2014)}]{petersen2014chiral}%
  \BibitemOpen
  \bibfield  {author} {\bibinfo {author} {\bibnamefont {Petersen},
  \bibfnamefont {J.}}, \bibinfo {author} {\bibnamefont {Volz}, \bibfnamefont
  {J.}}, \ and\ \bibinfo {author} {\bibnamefont {Rauschenbeutel}, \bibfnamefont
  {A.}},\ }\href@noop {} {\bibfield  {journal} {\bibinfo  {journal} {Science}\
  }\textbf {\bibinfo {volume} {346}},\ \bibinfo {pages} {67} (\bibinfo {year}
  {2014})}\BibitemShut {NoStop}%
\bibitem [{\citenamefont {Plotnik}\ \emph {et~al.}(2014)\citenamefont
  {Plotnik}, \citenamefont {Rechtsman}, \citenamefont {Song}, \citenamefont
  {Heinrich}, \citenamefont {Zeuner}, \citenamefont {Nolte}, \citenamefont
  {Lumer}, \citenamefont {Malkova}, \citenamefont {Xu}, \citenamefont {Szameit}
  \emph {et~al.}}]{rechtsman2012observation}%
  \BibitemOpen
  \bibfield  {author} {\bibinfo {author} {\bibnamefont {Plotnik}, \bibfnamefont
  {Y.}}, \bibinfo {author} {\bibnamefont {Rechtsman}, \bibfnamefont {M.~C.}},
  \bibinfo {author} {\bibnamefont {Song}, \bibfnamefont {D.}}, \bibinfo
  {author} {\bibnamefont {Heinrich}, \bibfnamefont {M.}}, \bibinfo {author}
  {\bibnamefont {Zeuner}, \bibfnamefont {J.~M.}}, \bibinfo {author}
  {\bibnamefont {Nolte}, \bibfnamefont {S.}}, \bibinfo {author} {\bibnamefont
  {Lumer}, \bibfnamefont {Y.}}, \bibinfo {author} {\bibnamefont {Malkova},
  \bibfnamefont {N.}}, \bibinfo {author} {\bibnamefont {Xu}, \bibfnamefont
  {J.}}, \bibinfo {author} {\bibnamefont {Szameit}, \bibfnamefont {A.}},  \emph
  {et~al.},\ }\href@noop {} {\bibfield  {journal} {\bibinfo  {journal} {Nature
  materials}\ }\textbf {\bibinfo {volume} {13}},\ \bibinfo {pages} {57}
  (\bibinfo {year} {2014})}\BibitemShut {NoStop}%
\bibitem [{\citenamefont {Pregnolato}\ \emph {et~al.}(2020)\citenamefont
  {Pregnolato}, \citenamefont {Chu}, \citenamefont {Schr{\"o}der},
  \citenamefont {Schott}, \citenamefont {Wieck}, \citenamefont {Ludwig},
  \citenamefont {Lodahl},\ and\ \citenamefont
  {Rotenberg}}]{pregnolato2020deterministic}%
  \BibitemOpen
  \bibfield  {author} {\bibinfo {author} {\bibnamefont {Pregnolato},
  \bibfnamefont {T.}}, \bibinfo {author} {\bibnamefont {Chu}, \bibfnamefont
  {X.-L.}}, \bibinfo {author} {\bibnamefont {Schr{\"o}der}, \bibfnamefont
  {T.}}, \bibinfo {author} {\bibnamefont {Schott}, \bibfnamefont {R.}},
  \bibinfo {author} {\bibnamefont {Wieck}, \bibfnamefont {A.~D.}}, \bibinfo
  {author} {\bibnamefont {Ludwig}, \bibfnamefont {A.}}, \bibinfo {author}
  {\bibnamefont {Lodahl}, \bibfnamefont {P.}}, \ and\ \bibinfo {author}
  {\bibnamefont {Rotenberg}, \bibfnamefont {N.}},\ }\href@noop {} {\bibfield
  {journal} {\bibinfo  {journal} {APL Photonics}\ }\textbf {\bibinfo {volume}
  {5}},\ \bibinfo {pages} {086101} (\bibinfo {year} {2020})}\BibitemShut
  {NoStop}%
\bibitem [{\citenamefont {Purcell}(1995)}]{purcell1995spontaneous}%
  \BibitemOpen
  \bibfield  {author} {\bibinfo {author} {\bibnamefont {Purcell}, \bibfnamefont
  {E.~M.}},\ }in\ \href@noop {} {\emph {\bibinfo {booktitle} {Confined
  Electrons and Photons}}}\ (\bibinfo  {publisher} {Springer},\ \bibinfo {year}
  {1995})\ pp.\ \bibinfo {pages} {839--839}\BibitemShut {NoStop}%
\bibitem [{\citenamefont {Qi}\ and\ \citenamefont
  {Zhang}(2011)}]{RevModPhys.83.1057}%
  \BibitemOpen
  \bibfield  {author} {\bibinfo {author} {\bibnamefont {Qi}, \bibfnamefont
  {X.-L.}}\ and\ \bibinfo {author} {\bibnamefont {Zhang}, \bibfnamefont
  {S.-C.}},\ }\href {\doibase 10.1103/RevModPhys.83.1057} {\bibfield  {journal}
  {\bibinfo  {journal} {Rev. Mod. Phys.}\ }\textbf {\bibinfo {volume} {83}},\
  \bibinfo {pages} {1057} (\bibinfo {year} {2011})}\BibitemShut {NoStop}%
\bibitem [{\citenamefont {Raghu}\ and\ \citenamefont
  {Haldane}(2008)}]{raghu2008analogs}%
  \BibitemOpen
  \bibfield  {author} {\bibinfo {author} {\bibnamefont {Raghu}, \bibfnamefont
  {S.}}\ and\ \bibinfo {author} {\bibnamefont {Haldane}, \bibfnamefont
  {F.~D.~M.}},\ }\href@noop {} {\bibfield  {journal} {\bibinfo  {journal}
  {Physical Review A}\ }\textbf {\bibinfo {volume} {78}},\ \bibinfo {pages}
  {033834} (\bibinfo {year} {2008})}\BibitemShut {NoStop}%
\bibitem [{\citenamefont {Rao}\ and\ \citenamefont
  {Hughes}(2007)}]{rao2007single}%
  \BibitemOpen
  \bibfield  {author} {\bibinfo {author} {\bibnamefont {Rao}, \bibfnamefont
  {V.~M.}}\ and\ \bibinfo {author} {\bibnamefont {Hughes}, \bibfnamefont
  {S.}},\ }\href@noop {} {\bibfield  {journal} {\bibinfo  {journal} {Physical
  review letters}\ }\textbf {\bibinfo {volume} {99}},\ \bibinfo {pages}
  {193901} (\bibinfo {year} {2007})}\BibitemShut {NoStop}%
\bibitem [{\citenamefont {Rechtsman}\ \emph {et~al.}(2013)\citenamefont
  {Rechtsman}, \citenamefont {Zeuner}, \citenamefont {Plotnik}, \citenamefont
  {Lumer}, \citenamefont {Podolsky}, \citenamefont {Dreisow}, \citenamefont
  {Nolte}, \citenamefont {Segev},\ and\ \citenamefont
  {Szameit}}]{rechtsman2013photonic}%
  \BibitemOpen
  \bibfield  {author} {\bibinfo {author} {\bibnamefont {Rechtsman},
  \bibfnamefont {M.~C.}}, \bibinfo {author} {\bibnamefont {Zeuner},
  \bibfnamefont {J.~M.}}, \bibinfo {author} {\bibnamefont {Plotnik},
  \bibfnamefont {Y.}}, \bibinfo {author} {\bibnamefont {Lumer}, \bibfnamefont
  {Y.}}, \bibinfo {author} {\bibnamefont {Podolsky}, \bibfnamefont {D.}},
  \bibinfo {author} {\bibnamefont {Dreisow}, \bibfnamefont {F.}}, \bibinfo
  {author} {\bibnamefont {Nolte}, \bibfnamefont {S.}}, \bibinfo {author}
  {\bibnamefont {Segev}, \bibfnamefont {M.}}, \ and\ \bibinfo {author}
  {\bibnamefont {Szameit}, \bibfnamefont {A.}},\ }\href@noop {} {\bibfield
  {journal} {\bibinfo  {journal} {Nature}\ }\textbf {\bibinfo {volume} {496}},\
  \bibinfo {pages} {196} (\bibinfo {year} {2013})}\BibitemShut {NoStop}%
\bibitem [{\citenamefont {Rodr{\'\i}guez-Fortu{\~n}o}\ \emph
  {et~al.}(2013)\citenamefont {Rodr{\'\i}guez-Fortu{\~n}o}, \citenamefont
  {Marino}, \citenamefont {Ginzburg}, \citenamefont {O’Connor}, \citenamefont
  {Mart{\'\i}nez}, \citenamefont {Wurtz},\ and\ \citenamefont
  {Zayats}}]{rodriguez2013near}%
  \BibitemOpen
  \bibfield  {author} {\bibinfo {author} {\bibnamefont
  {Rodr{\'\i}guez-Fortu{\~n}o}, \bibfnamefont {F.~J.}}, \bibinfo {author}
  {\bibnamefont {Marino}, \bibfnamefont {G.}}, \bibinfo {author} {\bibnamefont
  {Ginzburg}, \bibfnamefont {P.}}, \bibinfo {author} {\bibnamefont
  {O’Connor}, \bibfnamefont {D.}}, \bibinfo {author} {\bibnamefont
  {Mart{\'\i}nez}, \bibfnamefont {A.}}, \bibinfo {author} {\bibnamefont
  {Wurtz}, \bibfnamefont {G.~A.}}, \ and\ \bibinfo {author} {\bibnamefont
  {Zayats}, \bibfnamefont {A.~V.}},\ }\href@noop {} {\bibfield  {journal}
  {\bibinfo  {journal} {Science}\ }\textbf {\bibinfo {volume} {340}},\ \bibinfo
  {pages} {328} (\bibinfo {year} {2013})}\BibitemShut {NoStop}%
\bibitem [{\citenamefont {Romach}\ \emph {et~al.}(2015)\citenamefont {Romach},
  \citenamefont {Mueller}, \citenamefont {Unden}, \citenamefont {Rogers},
  \citenamefont {Isoda}, \citenamefont {Itoh}, \citenamefont {Markham},
  \citenamefont {Stacey}, \citenamefont {Meijer}, \citenamefont {Pezzagna},
  \citenamefont {Naydenov}, \citenamefont {McGuinness}, \citenamefont
  {Bar-Gill},\ and\ \citenamefont {Jelezko}}]{PhysRevLett.114.017601}%
  \BibitemOpen
  \bibfield  {author} {\bibinfo {author} {\bibnamefont {Romach}, \bibfnamefont
  {Y.}}, \bibinfo {author} {\bibnamefont {Mueller}, \bibfnamefont {C.}},
  \bibinfo {author} {\bibnamefont {Unden}, \bibfnamefont {T.}}, \bibinfo
  {author} {\bibnamefont {Rogers}, \bibfnamefont {L.~J.}}, \bibinfo {author}
  {\bibnamefont {Isoda}, \bibfnamefont {T.}}, \bibinfo {author} {\bibnamefont
  {Itoh}, \bibfnamefont {K.~M.}}, \bibinfo {author} {\bibnamefont {Markham},
  \bibfnamefont {M.}}, \bibinfo {author} {\bibnamefont {Stacey}, \bibfnamefont
  {A.}}, \bibinfo {author} {\bibnamefont {Meijer}, \bibfnamefont {J.}},
  \bibinfo {author} {\bibnamefont {Pezzagna}, \bibfnamefont {S.}}, \bibinfo
  {author} {\bibnamefont {Naydenov}, \bibfnamefont {B.}}, \bibinfo {author}
  {\bibnamefont {McGuinness}, \bibfnamefont {L.~P.}}, \bibinfo {author}
  {\bibnamefont {Bar-Gill}, \bibfnamefont {N.}}, \ and\ \bibinfo {author}
  {\bibnamefont {Jelezko}, \bibfnamefont {F.}},\ }\href {\doibase
  10.1103/PhysRevLett.114.017601} {\bibfield  {journal} {\bibinfo  {journal}
  {Phys. Rev. Lett.}\ }\textbf {\bibinfo {volume} {114}},\ \bibinfo {pages}
  {017601} (\bibinfo {year} {2015})}\BibitemShut {NoStop}%
\bibitem [{\citenamefont {Saba}\ \emph {et~al.}(2020)\citenamefont {Saba},
  \citenamefont {Wong}, \citenamefont {Elman}, \citenamefont {Oh},\ and\
  \citenamefont {Hess}}]{PhysRevB.101.054307}%
  \BibitemOpen
  \bibfield  {author} {\bibinfo {author} {\bibnamefont {Saba}, \bibfnamefont
  {M.}}, \bibinfo {author} {\bibnamefont {Wong}, \bibfnamefont {S.}}, \bibinfo
  {author} {\bibnamefont {Elman}, \bibfnamefont {M.}}, \bibinfo {author}
  {\bibnamefont {Oh}, \bibfnamefont {S.~S.}}, \ and\ \bibinfo {author}
  {\bibnamefont {Hess}, \bibfnamefont {O.}},\ }\href {\doibase
  10.1103/PhysRevB.101.054307} {\bibfield  {journal} {\bibinfo  {journal}
  {Phys. Rev. B}\ }\textbf {\bibinfo {volume} {101}},\ \bibinfo {pages}
  {054307} (\bibinfo {year} {2020})}\BibitemShut {NoStop}%
\bibitem [{\citenamefont {Sapienza}\ \emph {et~al.}(2010)\citenamefont
  {Sapienza}, \citenamefont {Thyrrestrup}, \citenamefont {Stobbe},
  \citenamefont {Garcia}, \citenamefont {Smolka},\ and\ \citenamefont
  {Lodahl}}]{sapienza2010cavity}%
  \BibitemOpen
  \bibfield  {author} {\bibinfo {author} {\bibnamefont {Sapienza},
  \bibfnamefont {L.}}, \bibinfo {author} {\bibnamefont {Thyrrestrup},
  \bibfnamefont {H.}}, \bibinfo {author} {\bibnamefont {Stobbe}, \bibfnamefont
  {S.}}, \bibinfo {author} {\bibnamefont {Garcia}, \bibfnamefont {P.~D.}},
  \bibinfo {author} {\bibnamefont {Smolka}, \bibfnamefont {S.}}, \ and\
  \bibinfo {author} {\bibnamefont {Lodahl}, \bibfnamefont {P.}},\ }\href@noop
  {} {\bibfield  {journal} {\bibinfo  {journal} {Science}\ }\textbf {\bibinfo
  {volume} {327}},\ \bibinfo {pages} {1352} (\bibinfo {year}
  {2010})}\BibitemShut {NoStop}%
\bibitem [{\citenamefont {Sauer}, \citenamefont {Vasco},\ and\ \citenamefont
  {Hughes}(2020)}]{PhysRevResearch.2.043109}%
  \BibitemOpen
  \bibfield  {author} {\bibinfo {author} {\bibnamefont {Sauer}, \bibfnamefont
  {E.}}, \bibinfo {author} {\bibnamefont {Vasco}, \bibfnamefont {J.~P.}}, \
  and\ \bibinfo {author} {\bibnamefont {Hughes}, \bibfnamefont {S.}},\ }\href
  {\doibase 10.1103/PhysRevResearch.2.043109} {\bibfield  {journal} {\bibinfo
  {journal} {Phys. Rev. Research}\ }\textbf {\bibinfo {volume} {2}},\ \bibinfo
  {pages} {043109} (\bibinfo {year} {2020})}\BibitemShut {NoStop}%
\bibitem [{\citenamefont {Sayrin}\ \emph {et~al.}(2015)\citenamefont {Sayrin},
  \citenamefont {Junge}, \citenamefont {Mitsch}, \citenamefont {Albrecht},
  \citenamefont {O’Shea}, \citenamefont {Schneeweiss}, \citenamefont {Volz},\
  and\ \citenamefont {Rauschenbeutel}}]{sayrin2015nanophotonic}%
  \BibitemOpen
  \bibfield  {author} {\bibinfo {author} {\bibnamefont {Sayrin}, \bibfnamefont
  {C.}}, \bibinfo {author} {\bibnamefont {Junge}, \bibfnamefont {C.}}, \bibinfo
  {author} {\bibnamefont {Mitsch}, \bibfnamefont {R.}}, \bibinfo {author}
  {\bibnamefont {Albrecht}, \bibfnamefont {B.}}, \bibinfo {author}
  {\bibnamefont {O’Shea}, \bibfnamefont {D.}}, \bibinfo {author}
  {\bibnamefont {Schneeweiss}, \bibfnamefont {P.}}, \bibinfo {author}
  {\bibnamefont {Volz}, \bibfnamefont {J.}}, \ and\ \bibinfo {author}
  {\bibnamefont {Rauschenbeutel}, \bibfnamefont {A.}},\ }\href@noop {}
  {\bibfield  {journal} {\bibinfo  {journal} {Physical Review X}\ }\textbf
  {\bibinfo {volume} {5}},\ \bibinfo {pages} {041036} (\bibinfo {year}
  {2015})}\BibitemShut {NoStop}%
\bibitem [{\citenamefont {Scheucher}\ \emph {et~al.}(2016)\citenamefont
  {Scheucher}, \citenamefont {Hilico}, \citenamefont {Will}, \citenamefont
  {Volz},\ and\ \citenamefont {Rauschenbeutel}}]{scheucher2016quantum}%
  \BibitemOpen
  \bibfield  {author} {\bibinfo {author} {\bibnamefont {Scheucher},
  \bibfnamefont {M.}}, \bibinfo {author} {\bibnamefont {Hilico}, \bibfnamefont
  {A.}}, \bibinfo {author} {\bibnamefont {Will}, \bibfnamefont {E.}}, \bibinfo
  {author} {\bibnamefont {Volz}, \bibfnamefont {J.}}, \ and\ \bibinfo {author}
  {\bibnamefont {Rauschenbeutel}, \bibfnamefont {A.}},\ }\href@noop {}
  {\bibfield  {journal} {\bibinfo  {journal} {Science}\ }\textbf {\bibinfo
  {volume} {354}},\ \bibinfo {pages} {1577} (\bibinfo {year}
  {2016})}\BibitemShut {NoStop}%
\bibitem [{\citenamefont {Schnauber}\ \emph {et~al.}(2018)\citenamefont
  {Schnauber}, \citenamefont {Schall}, \citenamefont {Bounouar}, \citenamefont
  {H\"ohne}, \citenamefont {Park}, \citenamefont {Ryu}, \citenamefont
  {Heindel}, \citenamefont {Burger}, \citenamefont {Song}, \citenamefont {Rodt}
  \emph {et~al.}}]{schnauber2018deterministic}%
  \BibitemOpen
  \bibfield  {author} {\bibinfo {author} {\bibnamefont {Schnauber},
  \bibfnamefont {P.}}, \bibinfo {author} {\bibnamefont {Schall}, \bibfnamefont
  {J.}}, \bibinfo {author} {\bibnamefont {Bounouar}, \bibfnamefont {S.}},
  \bibinfo {author} {\bibnamefont {H\"ohne}, \bibfnamefont {T.}}, \bibinfo
  {author} {\bibnamefont {Park}, \bibfnamefont {S.-I.}}, \bibinfo {author}
  {\bibnamefont {Ryu}, \bibfnamefont {G.-H.}}, \bibinfo {author} {\bibnamefont
  {Heindel}, \bibfnamefont {T.}}, \bibinfo {author} {\bibnamefont {Burger},
  \bibfnamefont {S.}}, \bibinfo {author} {\bibnamefont {Song}, \bibfnamefont
  {J.-D.}}, \bibinfo {author} {\bibnamefont {Rodt}, \bibfnamefont {S.}},  \emph
  {et~al.},\ }\href@noop {} {\bibfield  {journal} {\bibinfo  {journal} {Nano
  Letters}\ }\textbf {\bibinfo {volume} {18}},\ \bibinfo {pages} {2336}
  (\bibinfo {year} {2018})}\BibitemShut {NoStop}%
\bibitem [{\citenamefont {Schneeweiss}\ \emph {et~al.}(2017)\citenamefont
  {Schneeweiss}, \citenamefont {Zeiger}, \citenamefont {Hoinkes}, \citenamefont
  {Rauschenbeutel},\ and\ \citenamefont {Volz}}]{schneeweiss2017fiber}%
  \BibitemOpen
  \bibfield  {author} {\bibinfo {author} {\bibnamefont {Schneeweiss},
  \bibfnamefont {P.}}, \bibinfo {author} {\bibnamefont {Zeiger}, \bibfnamefont
  {S.}}, \bibinfo {author} {\bibnamefont {Hoinkes}, \bibfnamefont {T.}},
  \bibinfo {author} {\bibnamefont {Rauschenbeutel}, \bibfnamefont {A.}}, \ and\
  \bibinfo {author} {\bibnamefont {Volz}, \bibfnamefont {J.}},\ }\href@noop {}
  {\bibfield  {journal} {\bibinfo  {journal} {Optics letters}\ }\textbf
  {\bibinfo {volume} {42}},\ \bibinfo {pages} {85} (\bibinfo {year}
  {2017})}\BibitemShut {NoStop}%
\bibitem [{\citenamefont {Schulz}\ \emph {et~al.}(2010)\citenamefont {Schulz},
  \citenamefont {O’Faolain}, \citenamefont {Beggs}, \citenamefont {White},
  \citenamefont {Melloni},\ and\ \citenamefont
  {Krauss}}]{schulz2010dispersion}%
  \BibitemOpen
  \bibfield  {author} {\bibinfo {author} {\bibnamefont {Schulz}, \bibfnamefont
  {S.}}, \bibinfo {author} {\bibnamefont {O’Faolain}, \bibfnamefont {L.}},
  \bibinfo {author} {\bibnamefont {Beggs}, \bibfnamefont {D.~M.}}, \bibinfo
  {author} {\bibnamefont {White}, \bibfnamefont {T.~P.}}, \bibinfo {author}
  {\bibnamefont {Melloni}, \bibfnamefont {A.}}, \ and\ \bibinfo {author}
  {\bibnamefont {Krauss}, \bibfnamefont {T.~F.}},\ }\href@noop {} {\bibfield
  {journal} {\bibinfo  {journal} {Journal of Optics}\ }\textbf {\bibinfo
  {volume} {12}},\ \bibinfo {pages} {104004} (\bibinfo {year}
  {2010})}\BibitemShut {NoStop}%
\bibitem [{\citenamefont {Shalaev}, \citenamefont {Walasik},\ and\
  \citenamefont {Litchinitser}(2019)}]{shalaev2019optically}%
  \BibitemOpen
  \bibfield  {author} {\bibinfo {author} {\bibnamefont {Shalaev}, \bibfnamefont
  {M.~I.}}, \bibinfo {author} {\bibnamefont {Walasik}, \bibfnamefont {W.}}, \
  and\ \bibinfo {author} {\bibnamefont {Litchinitser}, \bibfnamefont {N.~M.}},\
  }\href@noop {} {\bibfield  {journal} {\bibinfo  {journal} {Optica}\ }\textbf
  {\bibinfo {volume} {6}},\ \bibinfo {pages} {839} (\bibinfo {year}
  {2019})}\BibitemShut {NoStop}%
\bibitem [{\citenamefont {Shalaev}\ \emph {et~al.}(2019)\citenamefont
  {Shalaev}, \citenamefont {Walasik}, \citenamefont {Tsukernik}, \citenamefont
  {Xu},\ and\ \citenamefont {Litchinitser}}]{shalaev2019robust}%
  \BibitemOpen
  \bibfield  {author} {\bibinfo {author} {\bibnamefont {Shalaev}, \bibfnamefont
  {M.~I.}}, \bibinfo {author} {\bibnamefont {Walasik}, \bibfnamefont {W.}},
  \bibinfo {author} {\bibnamefont {Tsukernik}, \bibfnamefont {A.}}, \bibinfo
  {author} {\bibnamefont {Xu}, \bibfnamefont {Y.}}, \ and\ \bibinfo {author}
  {\bibnamefont {Litchinitser}, \bibfnamefont {N.~M.}},\ }\href@noop {}
  {\bibfield  {journal} {\bibinfo  {journal} {Nature nanotechnology}\ }\textbf
  {\bibinfo {volume} {14}},\ \bibinfo {pages} {31} (\bibinfo {year}
  {2019})}\BibitemShut {NoStop}%
\bibitem [{\citenamefont {Shomroni}\ \emph {et~al.}(2014)\citenamefont
  {Shomroni}, \citenamefont {Rosenblum}, \citenamefont {Lovsky}, \citenamefont
  {Bechler}, \citenamefont {Guendelman},\ and\ \citenamefont
  {Dayan}}]{shomroni2014all}%
  \BibitemOpen
  \bibfield  {author} {\bibinfo {author} {\bibnamefont {Shomroni},
  \bibfnamefont {I.}}, \bibinfo {author} {\bibnamefont {Rosenblum},
  \bibfnamefont {S.}}, \bibinfo {author} {\bibnamefont {Lovsky}, \bibfnamefont
  {Y.}}, \bibinfo {author} {\bibnamefont {Bechler}, \bibfnamefont {O.}},
  \bibinfo {author} {\bibnamefont {Guendelman}, \bibfnamefont {G.}}, \ and\
  \bibinfo {author} {\bibnamefont {Dayan}, \bibfnamefont {B.}},\ }\href@noop {}
  {\bibfield  {journal} {\bibinfo  {journal} {Science}\ }\textbf {\bibinfo
  {volume} {345}},\ \bibinfo {pages} {903} (\bibinfo {year}
  {2014})}\BibitemShut {NoStop}%
\bibitem [{\citenamefont {Skirlo}, \citenamefont {Lu},\ and\ \citenamefont
  {Soljac}(2014)}]{PhysRevLett.113.113904}%
  \BibitemOpen
  \bibfield  {author} {\bibinfo {author} {\bibnamefont {Skirlo}, \bibfnamefont
  {S.~A.}}, \bibinfo {author} {\bibnamefont {Lu}, \bibfnamefont {L.}}, \ and\
  \bibinfo {author} {\bibnamefont {Soljac}, \bibfnamefont {M.}},\ }\href
  {\doibase 10.1103/PhysRevLett.113.113904} {\bibfield  {journal} {\bibinfo
  {journal} {Phys. Rev. Lett.}\ }\textbf {\bibinfo {volume} {113}},\ \bibinfo
  {pages} {113904} (\bibinfo {year} {2014})}\BibitemShut {NoStop}%
\bibitem [{\citenamefont {Skorobogatiy}, \citenamefont {B{\'e}gin},\ and\
  \citenamefont {Talneau}(2005)}]{skorobogatiy2005statistical}%
  \BibitemOpen
  \bibfield  {author} {\bibinfo {author} {\bibnamefont {Skorobogatiy},
  \bibfnamefont {M.}}, \bibinfo {author} {\bibnamefont {B{\'e}gin},
  \bibfnamefont {G.}}, \ and\ \bibinfo {author} {\bibnamefont {Talneau},
  \bibfnamefont {A.}},\ }\href@noop {} {\bibfield  {journal} {\bibinfo
  {journal} {Optics express}\ }\textbf {\bibinfo {volume} {13}},\ \bibinfo
  {pages} {2487} (\bibinfo {year} {2005})}\BibitemShut {NoStop}%
\bibitem [{\citenamefont {S{\"o}llner}\ \emph {et~al.}(2015)\citenamefont
  {S{\"o}llner}, \citenamefont {Mahmoodian}, \citenamefont {Hansen},
  \citenamefont {Midolo}, \citenamefont {Javadi}, \citenamefont
  {Kir{\v{s}}ansk{\.e}}, \citenamefont {Pregnolato}, \citenamefont {El-Ella},
  \citenamefont {Lee}, \citenamefont {Song} \emph
  {et~al.}}]{sollner2015deterministic}%
  \BibitemOpen
  \bibfield  {author} {\bibinfo {author} {\bibnamefont {S{\"o}llner},
  \bibfnamefont {I.}}, \bibinfo {author} {\bibnamefont {Mahmoodian},
  \bibfnamefont {S.}}, \bibinfo {author} {\bibnamefont {Hansen}, \bibfnamefont
  {S.~L.}}, \bibinfo {author} {\bibnamefont {Midolo}, \bibfnamefont {L.}},
  \bibinfo {author} {\bibnamefont {Javadi}, \bibfnamefont {A.}}, \bibinfo
  {author} {\bibnamefont {Kir{\v{s}}ansk{\.e}}, \bibfnamefont {G.}}, \bibinfo
  {author} {\bibnamefont {Pregnolato}, \bibfnamefont {T.}}, \bibinfo {author}
  {\bibnamefont {El-Ella}, \bibfnamefont {H.}}, \bibinfo {author} {\bibnamefont
  {Lee}, \bibfnamefont {E.~H.}}, \bibinfo {author} {\bibnamefont {Song},
  \bibfnamefont {J.~D.}},  \emph {et~al.},\ }\href@noop {} {\bibfield
  {journal} {\bibinfo  {journal} {Nature nanotechnology}\ }\textbf {\bibinfo
  {volume} {10}},\ \bibinfo {pages} {775} (\bibinfo {year} {2015})}\BibitemShut
  {NoStop}%
\bibitem [{\citenamefont {Vasco}\ and\ \citenamefont
  {Hughes}(2017)}]{PhysRevB.95.224202}%
  \BibitemOpen
  \bibfield  {author} {\bibinfo {author} {\bibnamefont {Vasco}, \bibfnamefont
  {J.~P.}}\ and\ \bibinfo {author} {\bibnamefont {Hughes}, \bibfnamefont
  {S.}},\ }\href {\doibase 10.1103/PhysRevB.95.224202} {\bibfield  {journal}
  {\bibinfo  {journal} {Phys. Rev. B}\ }\textbf {\bibinfo {volume} {95}},\
  \bibinfo {pages} {224202} (\bibinfo {year} {2017})}\BibitemShut {NoStop}%
\bibitem [{\citenamefont {Volz}\ \emph {et~al.}(2012)\citenamefont {Volz},
  \citenamefont {Reinhard}, \citenamefont {Winger}, \citenamefont {Badolato},
  \citenamefont {Hennessy}, \citenamefont {Hu},\ and\ \citenamefont
  {Imamo{\u{g}}lu}}]{volz2012ultrafast}%
  \BibitemOpen
  \bibfield  {author} {\bibinfo {author} {\bibnamefont {Volz}, \bibfnamefont
  {T.}}, \bibinfo {author} {\bibnamefont {Reinhard}, \bibfnamefont {A.}},
  \bibinfo {author} {\bibnamefont {Winger}, \bibfnamefont {M.}}, \bibinfo
  {author} {\bibnamefont {Badolato}, \bibfnamefont {A.}}, \bibinfo {author}
  {\bibnamefont {Hennessy}, \bibfnamefont {K.~J.}}, \bibinfo {author}
  {\bibnamefont {Hu}, \bibfnamefont {E.~L.}}, \ and\ \bibinfo {author}
  {\bibnamefont {Imamo{\u{g}}lu}, \bibfnamefont {A.}},\ }\href@noop {}
  {\bibfield  {journal} {\bibinfo  {journal} {Nature Photonics}\ }\textbf
  {\bibinfo {volume} {6}},\ \bibinfo {pages} {605} (\bibinfo {year}
  {2012})}\BibitemShut {NoStop}%
\bibitem [{\citenamefont {Wang}\ \emph {et~al.}(2008)\citenamefont {Wang},
  \citenamefont {Mazoyer}, \citenamefont {Hugonin},\ and\ \citenamefont
  {Lalanne}}]{wang2008backscattering}%
  \BibitemOpen
  \bibfield  {author} {\bibinfo {author} {\bibnamefont {Wang}, \bibfnamefont
  {B.}}, \bibinfo {author} {\bibnamefont {Mazoyer}, \bibfnamefont {S.}},
  \bibinfo {author} {\bibnamefont {Hugonin}, \bibfnamefont {J.-P.}}, \ and\
  \bibinfo {author} {\bibnamefont {Lalanne}, \bibfnamefont {P.}},\ }\href@noop
  {} {\bibfield  {journal} {\bibinfo  {journal} {Physical Review B}\ }\textbf
  {\bibinfo {volume} {78}},\ \bibinfo {pages} {245108} (\bibinfo {year}
  {2008})}\BibitemShut {NoStop}%
\bibitem [{\citenamefont {Wang}\ \emph {et~al.}(2009)\citenamefont {Wang},
  \citenamefont {Chong}, \citenamefont {Joannopoulos},\ and\ \citenamefont
  {Soljac}}]{wang2009observation}%
  \BibitemOpen
  \bibfield  {author} {\bibinfo {author} {\bibnamefont {Wang}, \bibfnamefont
  {Z.}}, \bibinfo {author} {\bibnamefont {Chong}, \bibfnamefont {Y.}}, \bibinfo
  {author} {\bibnamefont {Joannopoulos}, \bibfnamefont {J.~D.}}, \ and\
  \bibinfo {author} {\bibnamefont {Soljac}, \bibfnamefont {M.}},\ }\href@noop
  {} {\bibfield  {journal} {\bibinfo  {journal} {Nature}\ }\textbf {\bibinfo
  {volume} {461}},\ \bibinfo {pages} {772} (\bibinfo {year}
  {2009})}\BibitemShut {NoStop}%
\bibitem [{\citenamefont {Xia}\ \emph {et~al.}(2014)\citenamefont {Xia},
  \citenamefont {Lu}, \citenamefont {Lin}, \citenamefont {Cheng}, \citenamefont
  {Niu}, \citenamefont {Gong}, \citenamefont {Twamley} \emph
  {et~al.}}]{xia2014reversible}%
  \BibitemOpen
  \bibfield  {author} {\bibinfo {author} {\bibnamefont {Xia}, \bibfnamefont
  {K.}}, \bibinfo {author} {\bibnamefont {Lu}, \bibfnamefont {G.}}, \bibinfo
  {author} {\bibnamefont {Lin}, \bibfnamefont {G.}}, \bibinfo {author}
  {\bibnamefont {Cheng}, \bibfnamefont {Y.}}, \bibinfo {author} {\bibnamefont
  {Niu}, \bibfnamefont {Y.}}, \bibinfo {author} {\bibnamefont {Gong},
  \bibfnamefont {S.}}, \bibinfo {author} {\bibnamefont {Twamley}, \bibfnamefont
  {J.}},  \emph {et~al.},\ }\href@noop {} {\bibfield  {journal} {\bibinfo
  {journal} {Physical Review A}\ }\textbf {\bibinfo {volume} {90}},\ \bibinfo
  {pages} {043802} (\bibinfo {year} {2014})}\BibitemShut {NoStop}%
\bibitem [{\citenamefont {Yamaguchi}\ \emph {et~al.}(2019)\citenamefont
  {Yamaguchi}, \citenamefont {Ota}, \citenamefont {Katsumi}, \citenamefont
  {Watanabe}, \citenamefont {Ishida}, \citenamefont {Osada}, \citenamefont
  {Arakawa},\ and\ \citenamefont {Iwamoto}}]{yamaguchi2019gaas}%
  \BibitemOpen
  \bibfield  {author} {\bibinfo {author} {\bibnamefont {Yamaguchi},
  \bibfnamefont {T.}}, \bibinfo {author} {\bibnamefont {Ota}, \bibfnamefont
  {Y.}}, \bibinfo {author} {\bibnamefont {Katsumi}, \bibfnamefont {R.}},
  \bibinfo {author} {\bibnamefont {Watanabe}, \bibfnamefont {K.}}, \bibinfo
  {author} {\bibnamefont {Ishida}, \bibfnamefont {S.}}, \bibinfo {author}
  {\bibnamefont {Osada}, \bibfnamefont {A.}}, \bibinfo {author} {\bibnamefont
  {Arakawa}, \bibfnamefont {Y.}}, \ and\ \bibinfo {author} {\bibnamefont
  {Iwamoto}, \bibfnamefont {S.}},\ }\href@noop {} {\bibfield  {journal}
  {\bibinfo  {journal} {Applied Physics Express}\ }\textbf {\bibinfo {volume}
  {12}},\ \bibinfo {pages} {062005} (\bibinfo {year} {2019})}\BibitemShut
  {NoStop}%
\bibitem [{\citenamefont {Yang}, \citenamefont {Jiang},\ and\ \citenamefont
  {Hang}(2018)}]{yang2018topological}%
  \BibitemOpen
  \bibfield  {author} {\bibinfo {author} {\bibnamefont {Yang}, \bibfnamefont
  {Y.}}, \bibinfo {author} {\bibnamefont {Jiang}, \bibfnamefont {H.}}, \ and\
  \bibinfo {author} {\bibnamefont {Hang}, \bibfnamefont {Z.~H.}},\ }\href@noop
  {} {\bibfield  {journal} {\bibinfo  {journal} {Scientific reports}\ }\textbf
  {\bibinfo {volume} {8}},\ \bibinfo {pages} {1} (\bibinfo {year}
  {2018})}\BibitemShut {NoStop}%
\bibitem [{\citenamefont {Young}\ \emph
  {et~al.}(2015{\natexlab{a}})\citenamefont {Young}, \citenamefont {Thijssen},
  \citenamefont {Beggs}, \citenamefont {Androvitsaneas}, \citenamefont
  {Kuipers}, \citenamefont {Rarity}, \citenamefont {Hughes},\ and\
  \citenamefont {Oulton}}]{young2015polarization}%
  \BibitemOpen
  \bibfield  {author} {\bibinfo {author} {\bibnamefont {Young}, \bibfnamefont
  {A.~B.}}, \bibinfo {author} {\bibnamefont {Thijssen}, \bibfnamefont {A.}},
  \bibinfo {author} {\bibnamefont {Beggs}, \bibfnamefont {D.~M.}}, \bibinfo
  {author} {\bibnamefont {Androvitsaneas}, \bibfnamefont {P.}}, \bibinfo
  {author} {\bibnamefont {Kuipers}, \bibfnamefont {L.}}, \bibinfo {author}
  {\bibnamefont {Rarity}, \bibfnamefont {J.~G.}}, \bibinfo {author}
  {\bibnamefont {Hughes}, \bibfnamefont {S.}}, \ and\ \bibinfo {author}
  {\bibnamefont {Oulton}, \bibfnamefont {R.}},\ }\href@noop {} {\bibfield
  {journal} {\bibinfo  {journal} {Physical review letters}\ }\textbf {\bibinfo
  {volume} {115}},\ \bibinfo {pages} {153901} (\bibinfo {year}
  {2015}{\natexlab{a}})}\BibitemShut {NoStop}%
\bibitem [{\citenamefont {Young}\ \emph
  {et~al.}(2015{\natexlab{b}})\citenamefont {Young}, \citenamefont {Thijssen},
  \citenamefont {Beggs}, \citenamefont {Androvitsaneas}, \citenamefont
  {Kuipers}, \citenamefont {Rarity}, \citenamefont {Hughes},\ and\
  \citenamefont {Oulton}}]{PhysRevLett.115.153901}%
  \BibitemOpen
  \bibfield  {author} {\bibinfo {author} {\bibnamefont {Young}, \bibfnamefont
  {A.~B.}}, \bibinfo {author} {\bibnamefont {Thijssen}, \bibfnamefont
  {A.~C.~T.}}, \bibinfo {author} {\bibnamefont {Beggs}, \bibfnamefont {D.~M.}},
  \bibinfo {author} {\bibnamefont {Androvitsaneas}, \bibfnamefont {P.}},
  \bibinfo {author} {\bibnamefont {Kuipers}, \bibfnamefont {L.}}, \bibinfo
  {author} {\bibnamefont {Rarity}, \bibfnamefont {J.~G.}}, \bibinfo {author}
  {\bibnamefont {Hughes}, \bibfnamefont {S.}}, \ and\ \bibinfo {author}
  {\bibnamefont {Oulton}, \bibfnamefont {R.}},\ }\href {\doibase
  10.1103/PhysRevLett.115.153901} {\bibfield  {journal} {\bibinfo  {journal}
  {Phys. Rev. Lett.}\ }\textbf {\bibinfo {volume} {115}},\ \bibinfo {pages}
  {153901} (\bibinfo {year} {2015}{\natexlab{b}})}\BibitemShut {NoStop}%
\end{thebibliography}%

\end{document}